\documentclass[aps,prd,reprint,nofootinbib,superscriptaddress,showkeys]{revtex4-2}

\usepackage{amssymb}   
\usepackage{mathtools} 
\usepackage{hyperref}
\hypersetup{colorlinks=true,linkcolor=blue,urlcolor=blue,citecolor=blue}
\usepackage{accents}
\usepackage{tensor}
\usepackage[cal=boondox]{mathalfa}
\usepackage{lipsum}
\usepackage{relsize}
\usepackage{color}
\usepackage{comment}
\usepackage{nameref}
\newcommand{\intprod}{\mbox{$\;
		\put(0,0){\line(1,0){.9}}\put(.9,0){\line(0,1){1.6}} \; \, \, $}}

\usepackage[T1]{fontenc}

\begin{document}

	\title{Off-Shell Noether Currents and Potentials for First-Order General Relativity}

	\author{Merced Montesinos\href{https://orcid.org/0000-0002-4936-9170} {\includegraphics[scale=0.05]{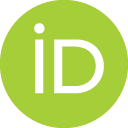}}}
	\email{merced@fis.cinvestav.mx}
	\author{Diego Gonzalez\href{https://orcid.org/0000-0002-0206-7378} {\includegraphics[scale=0.05]{ORCIDiD_icon128x128.png}}}
	\email{dgonzalez@fis.cinvestav.mx}
	\author{Rodrigo Romero\href{https://orcid.org/0000-0002-5529-6991} {\includegraphics[scale=0.05]{ORCIDiD_icon128x128.png}}}
	\email{rromero@fis.cinvestav.mx}
	
	\affiliation{Departamento de F\'{i}sica, Cinvestav, Avenida Instituto Polit\'{e}cnico Nacional 2508,\\
		San Pedro Zacatenco, 07360 Gustavo A. Madero, Ciudad de M\'exico, Mexico}
	
	\author{Mariano Celada\href{https://orcid.org/0000-0002-3519-4736} {\includegraphics[scale=0.05]{ORCIDiD_icon128x128.png}}}
	\email[]{mcelada@matmor.unam.mx}
	\affiliation{Centro de Ciencias Matem\'{a}ticas, Universidad Nacional Aut\'{o}noma de M\'{e}xico,\\
	UNAM-Campus Morelia, Apartado Postal 61-3, Morelia, Michoac\'{a}n 58090, Mexico}
	
	\date{\today}
	
	\begin{abstract}
		We report off-shell Noether currents obtained from off-shell Noether potentials for first-order general relativity described by $n$-dimensional Palatini and Holst Lagrangians including the cosmological constant. These off-shell currents and potentials are achieved by using the corresponding Lagrangian and the off-shell Noether identities satisfied by diffeomorphisms generated by arbitrary vector fields, local $SO(n)$ or $SO(n-1,1)$ transformations, `improved diffeomorphisms', and the `generalization of local translations' of the orthonormal frame and the connection. A remarkable aspect of our approach is that we do {\it not} use Noether's theorem in its direct form. By construction, the currents are off-shell conserved and lead naturally to the definition of off-shell Noether charges. We also study what we call the `half off-shell' case for both Palatini and Holst Lagrangians. In particular, we find that the resulting diffeomorphism and local $SO(3,1)$ or $SO(4)$ off-shell Noether currents and potentials for the Holst Lagrangian generically depend on the Immirzi parameter, which holds even in the `half off-shell' and on-shell cases. We also study Killing vector fields in the `half off-shell' and on-shell cases. The current theoretical framework is illustrated for the `half off-shell' case in static spherically symmetric and Friedmann--Lemaitre--Robertson--Walker spacetimes in four dimensions.
	\end{abstract}
	
	\keywords{Noether potentials; Noether currents; Noether charges; Palatini Lagrangian; Holst Lagrangian} 
	
	\maketitle
	\onecolumngrid
	
\section{Introduction}

Some of the most fundamental results in mathematical physics are Noether's theorems~\cite{Noether,Bessel-Hagen,EmmyNoether}, which establish a deep connection between infinitesimal symmetries of a variational principle and conservation laws. There are two Noether's theorems: the first one dealing with global (or rigid) symmetries and the second one concerning local \linebreak(or gauge) symmetries. These theorems have forged and shaped the modern view of theoretical physics, furnishing a vast amount of applications in many areas of physics. 

Noether's theorems provide powerful tools to calculate conserved currents and charges of physical systems. This feature has been highly exploited by the gravitational community and is an active ingredient of the modern research in general relativity and other alternative theories of gravity \cite{Wald1,Julia_1998,Fatibene,Wald3,Fatibene2,BARNICH20023,Barnich_2003}. For instance, the definition of energy in generally covariant systems is a rather delicate issue \cite{Deser_2019} and Noether's theorem plays a central role in addressing it.

To obtain conserved charges associated to spacetime diffeomorphisms in gravitational systems, Noether's second theorem is implemented in its direct fashion (its converse also holds), leading to the construction of an associated Noether current that is conserved on-shell \cite{Wald1,Wald3,BARNICH20023}; that is, when the equations of motion are satisfied. This is the usual viewpoint taken and one might wonder whether it is really necessary to work on-shell to obtain such conservation laws. After all, in any gauge theory,  there exist Noether currents that are identically off-shell conserved \cite{Julia_1998}, thus leading to the definition of off-shell potentials and charges. 

Some years ago, off-shell Noether currents and potentials were introduced to define quasi-local charges in any theory of metric gravity invariant under diffeomorphisms \cite{Kim1}. Later on, an analogous proposal was put forward for covariant gravity theories within the first-order formalism \cite{adami2016,ADAMI20191}, with again off-shell Noether currents and potentials playing an essential role, although only the case of a combined Lorentz-diffeomorphism symmetry is considered. 

It is well-known that, within the first-order formalism, general relativity can be described by either the Palatini action or the Holst action \cite{Holst}, the latter   being the starting point of the loop approach to quantum gravity in its canonical and covariant versions~\cite{Rovebook,Thiebook,perez2013,rovelli2014covariant}. In particular, the Holst action contains the so-called Immirzi or Barbero--Immirzi parameter~\cite{Barbero, Immirzi}, which plays no role when the equations of motion are satisfied, but on which the theory strongly depends when away from them (off-shell). In fact, this parameter affects the way in which fermions couple to gravity \cite{Freidel2005,Mercuri2006,Perez-Rovelli-2006,BojowaldDas2008} and manifests itself in the spectra of geometric quantum operators \cite{RoveSmolinNucP442,AshtLewacqg14,Thiebook} and in the black hole \mbox{entropy \cite{RovPRLt.77.3288,AshBaezcqg105,Meisscqg21,Agulloetcprl100,EngleNuiPRL105}} derived within the loop framework. Then, given the relevance of the Immirzi parameter at both the classical and quantum levels, it is important to understand how this parameter may contribute to off-shell Noether currents and potentials.

To expand our horizons on the off-shell effects of the Immirzi parameter, in this paper, by using a new theoretical framework, we find off-shell Noether currents and potentials for general relativity with a cosmological constant in the first-order formalism for both the Palatini Lagrangian in $n$ dimensions and the Holst Lagrangian in four dimensions. The advantage of our approach is that it is carried out completely {\it off-shell}, and its novelty is that it takes advantage of the off-shell Noether identities arising from both Lagrangians and avoids the use of Noether's second theorem in its direct version. We report these off-shell Noether currents and potentials for infinitesimal diffeomorphisms generated by arbitrary vector fields, local $SO(n)$ or $SO(n-1,1)$ transformations, `improved diffeomorphisms', and the so-called `generalization of local translations' of the orthonormal frame and the connection \cite{Montesinos1}. The resulting off-shell Noether current and potential for diffeomorphisms can be regarded as the first-order version of the off-shell Noether current and potential in the metric second-order formalism reported in \cite{Kim1}. Remarkably, our results reveal that the Immirzi parameter affects, in a non-trivial way, the definition of the off-shell Noether currents and potentials associated to diffeomorphisms and local $SO(3,1)$ or $SO(4)$ transformations for the Holst Lagrangian. We also find that the currents for both `improved diffeomorphisms' and the `generalization of local translations' identically vanish. Nevertheless, we also show that from these symmetries we can obtain the off-shell current and potential for diffeomorphisms for both Lagrangians. 

Additionally, we consider the particular case of diffeomorphisms generated by Killing vector fields and determine their corresponding off-shell Noether currents and potentials. This leads to the introduction of an effective gauge transformation, and we report its corresponding off-shell Noether currents and potentials. We also work `half off-shell' (in a sense made precise in Section \ref{HOS}) and obtain general expressions for the Noether currents and potentials for both Palatini and Holst Lagrangians. We find that in the `half off-shell' case the resulting diffeomorphism and $SO(3,1)$ or $SO(4)$ Noether currents and potentials for the Holst Lagrangian still generically depend on the Immirzi parameter, even though the ones for the second-order Lagrangian (Einstein--Hilbert Lagrangian in terms of the tetrad) are independent of it. This implies that the Immirzi parameter is going to be present in these currents and potentials even `on-shell'. Furthermore, even though the Noether potential for the effective gauge transformation for the Holst Lagrangian in the `half off-shell' case still depends on the Immirzi parameter, the `half off-shell' current is independent of it. Finally, we illustrate the current theoretical framework in four-dimensional static spherically symmetric and Friedmann--Lemaitre--Robertson--Walker (FLRW) spacetimes  and report the explicit expressions for the `half off-shell' Noether currents and potentials, which turn out to depend on the Immirzi parameter as expected. 

We follow the notation and conventions of   \citet{Montesinos1}. Let $M$ be an $n$-dimensional Lorentzian or Riemannian manifold. In the first-order formalism, the fundamental variables are an orthonormal frame of $1$-forms $e^I$ and a connection 1-form $\omega^I{}_J$ compatible with the metric $(\eta_{IJ}) =\mbox{diag} (\sigma,1, \ldots, 1)$, $d \eta_{IJ} - \omega^K{}_I \eta_{KJ} - \omega^K{}_J \eta_{IK}=0$, and therefore $\omega_{IJ}=-\omega_{JI}$ because frame indices $I,J,K, \ldots$, which take the values $0,1,\ldots,n-1$, are raised and lowered with $\eta_{IJ}$. For $\sigma=-1$   the frame rotation group is the Lorentz group $SO(n-1,1)$, whereas for $\sigma =1$ it is the rotation group $SO(n)$. The $SO(n-1,1)$ [or $SO(n)$] totally antisymmetric tensor $\epsilon_{I_1 \ldots I_n}$ is such that $\epsilon_{01 \ldots n-1}=1$. The symbols $\wedge$, $d$, and ${\mathcal L}_{\zeta}$ stand for the wedge product, exterior derivative, and the Lie derivative along the vector $\zeta$ of differential forms, respectively. Furthermore, $\intprod$ stands for the contraction of a vector field with a differential form \cite{Castillobook}, the volume form is given by $\eta = (1/n!) \epsilon_{I_1 \ldots I_n} e^{I_1} \wedge \cdots \wedge e^{I_n}$, $\star$ is the Hodge dual, and $D$ stands for the covariant derivative with respect to $\omega^I{}_J$. The antisymmetric part of tensors involving frame indices is defined by $t^{[IJ]} = \left (t^{IJ} - t^{JI} \right)/2$.

It is worth pointing out that in this paper we focus our attention on the Lagrangian $n$-form instead of the action principle, which in turn is defined as the integral of the Lagrangian over a determined spacetime region. Thus, the Lagrangian itself completely specifies the theory under consideration.

\section{Palatini Lagrangian}\label{Palatini}
First-order general relativity in $n$-dimensions with (or without) a cosmological constant $\Lambda$ can be described by the action principle constructed out of the $n$-dimensional Palatini Lagrangian 
\begin{equation}
	L_P = \kappa \left [ R^{IJ} \wedge \star (e_I \wedge e_J) - 2 \Lambda \eta \right ], \label{PalatiniL}
\end{equation}
where $R^I{}_J = d \omega^I{}_J + \omega^I{}_K \wedge \omega^K{}_J$ is the curvature of $\omega^I{}_J$. Because we do not consider matter fields in this paper, we could omit the constant $\kappa:=(16\pi G)^{-1}$ in the previous Lagrangian. However, we   keep it for dimensional reasons (so that the Lagrangian has dimensions of action).

A general variation of the Palatini Lagrangian under the corresponding variations of the frame $e^I$ and the connection $\omega^I{}_J$ takes the form
\begin{equation}
	\delta L_P = {\mathcal E}_I \wedge \delta e^I + {\mathcal E}_{IJ} \wedge \delta \omega^{IJ} + d \left [ \kappa \delta \omega^{IJ} \wedge \star \left ( e_I \wedge e_J \right ) \right ], \label{variation}
\end{equation}
where the variational derivatives ${\mathcal E}_I$ and ${\mathcal E}_{IJ}$ are given by
\begin{eqnarray}
	{\mathcal E}_I &=& \kappa (-1)^{n-1} \left [ \star \left ( e_I \wedge e_J \wedge e_K \right ) \wedge R^{JK} - 2 \Lambda \star e_I \right ], \label{dervarP1}\\
	{\mathcal E}_{IJ} &=& \kappa (-1)^{n-1} D \left [ \star \left ( e_I \wedge e_J \right ) \right ]. \label{dervarP2}
\end{eqnarray}

\subsection{Off-Shell Current and Potential for Diffeomorphisms}\label{Lied}

By handling the variational derivatives ${\mathcal E}_I$ and ${\mathcal E}_{IJ}$ given in \eqref{dervarP1} and \eqref{dervarP2}, we get the off-shell Noether identity \cite{Montesinos1}
\begin{equation} \label{NIdiff}
	{\mathcal E}_I \wedge {\mathcal L}_{\zeta} e^I + {\mathcal E}_{IJ} \wedge {\mathcal L}_{\zeta} \omega^{IJ} + d \left \{ (-1)^n \left [ (\zeta \intprod \omega^{IJ}) {\mathcal E}_{IJ} + (\zeta \intprod e^I ) {\mathcal E}_I \right ] \right \} =0, 
\end{equation}
satisfied by the change of $e^I$ and $\omega^I{}_J$ under an infinitesimal diffeomorphism generated by an arbitrary vector field $\zeta$ ({\it converse} of Noether's second theorem). 

By computing the variation \eqref{variation} for the change of $e^I$ and $\omega^I{}_J$ under an infinitesimal diffeomorphism generated by $\zeta$, we obtain
\begin{equation}
	\delta_{\zeta} L_P = {\mathcal E}_I \wedge {\mathcal L}_{\zeta} e^I + {\mathcal E}_{IJ} \wedge {\mathcal L}_{\zeta} \omega^{IJ} + d \left [ \kappa {\mathcal L}_{\zeta} \omega^{IJ} \wedge \star \left ( e_I \wedge e_J \right ) \right ]. \label{variationdiff}
\end{equation}
Using \eqref{NIdiff}, we rewrite the right-hand side of last expression as
\begin{equation}
	\delta_{\zeta} L_P = d \left \{ (-1)^{n-1} \left [ (\zeta \intprod \omega^{IJ}) {\mathcal E}_{IJ} + (\zeta \intprod e^I ) {\mathcal E}_I - \kappa \star \left ( e_I \wedge e_J \right ) \wedge {\mathcal L}_{\zeta} \omega^{IJ} \right ] \right \}, \label{vardiff}
\end{equation}
that is, as an exact form. It is remarkable that the terms inside the braces can be written as 
\begin{eqnarray}
	&&(-1)^{n-1} \left [ (\zeta \intprod \omega^{IJ}) {\mathcal E}_{IJ} + (\zeta \intprod e^I ) {\mathcal E}_I - \kappa \star \left ( e_I \wedge e_J \right ) \wedge {\mathcal L}_{\zeta} \omega^{IJ} \right ] \nonumber\\
	&&= d \left [ \kappa \left ( \zeta \intprod \omega^{IJ} \right ) \star \left ( e_I \wedge e_J \right ) \right ] 
	+ \zeta \intprod L_P. \label{crucial}
\end{eqnarray} 
The meaning of the off-shell identity \eqref{crucial} is better appreciated by noting that it has the form
\begin{equation}
	J_{\zeta} = d U_{\zeta}, \label{keyeq}
\end{equation}
where the off-shell current $J_{\zeta}$   is defined by 
\begin{equation}
	J_{\zeta} := - \zeta \intprod L_P + (-1)^{n-1} \left [ (\zeta \intprod \omega^{IJ}) {\mathcal E}_{IJ} + (\zeta \intprod e^I ) {\mathcal E}_I - \kappa \star \left ( e_I \wedge e_J \right ) \wedge {\mathcal L}_{\zeta} \omega^{IJ} \right ],\label{Jdiff}
\end{equation}
with corresponding off-shell Noether potential $U_{\zeta}$ defined by 
\begin{equation}
	U_{\zeta} := \kappa \left ( \zeta \intprod \omega^{IJ} \right ) \star \left ( e_I \wedge e_J \right ). \label{Udiff}
\end{equation}
It follows from \eqref{keyeq} that $J_{\zeta}$ is off-shell conserved 
\begin{equation}
	d J_{\zeta} =0.
\end{equation}
Note that the off-shell current \eqref{Jdiff} can be further simplified and it acquires the off-shell form
\begin{equation}
	J_{\zeta} = (-1)^{n-1} \left ( \zeta \intprod \omega^{IJ} \right ) {\mathcal E}_{IJ} + (-1)^n \kappa \star \left ( e_I \wedge e_J \right ) 
	\wedge D \left ( \zeta \intprod \omega^{IJ} \right ). \label{Jdiff2}
\end{equation}
This expression is remarkable because it involves neither ${\mathcal E}_I$ nor $L_P$, in contrast to \eqref{Jdiff}.

Here, we show  three things: First, we provide a new and systematic theoretical framework that allows us to define the {\it off-shell} Noether potential $U_{\zeta}$ given by \eqref{Udiff} and the {\it off-shell} Noether current $J_{\zeta}$ given by \eqref{Jdiff} associated to the diffeomorphism covariance of the Palatini Lagrangian \eqref{PalatiniL}. Second, we show  that $U_{\zeta}$ and $J_{\zeta}$ are related by \eqref{keyeq}. Third, we show  that $J_{\zeta}$ is {\it off-shell} conserved too. The whole procedure to achieve these three things is carried out {\it off-shell}, namely, without using the equations of motion. This is in contrast to the conventional approach found in literature (see, e.g.,  \cite{Jacobson}), where these three things are defined only on-shell. Therefore, we have generalized and extended these notions to the first-order formalism of general relativity described by the Palatini Lagrangian, and shown that it is {\it not} necessary to define these notions on-shell. Thus, it is correct to interpret these results coming from the Palatini Lagrangian as the first-order version of the off-shell Noether potential and current associated to diffeomorphisms for general relativity in the metric second-order formalism reported in \cite{Kim1}. Further, our approach is also very general in the sense that it holds for any arbitrary vector field $\zeta$. 

Let us emphasize that the off-shell aspect is just one of the important features of the theoretical framework developed in this paper. The second aspect is that Noether's theorem for gauge transformations (also called Noether's second theorem) was {\it not} used at all to get~\eqref{keyeq}. This is another key difference between the approach of this paper and previous ones~\cite{Wald1,Jacobson,Corichi2,oliveri2020}, i.e., we did {\it not} assume (nor use) that under an infinitesimal diffeomorphism generated by $\zeta$ the change of the action is 
\begin{equation}
	\delta_{\zeta} S_P [e,\omega] = \delta_{\zeta} \int_{\mathcal M} L_P = \int_{\mathcal M} {\mathcal L}_{\zeta} L_P =\int_{\mathcal M} d \left ( \zeta \intprod L\right), \label{change_action}
\end{equation} 
as is usually assumed when dealing with Noether's theorem for diffeomorphism transformations. In our approach, the relation \eqref{change_action} holds, of course, but it is {\it deduced} from the combination of \eqref{vardiff} and \eqref{crucial}.

\subsection{Off-Shell Current and Potential for Local $SO(n-1,1)$ or $SO(n)$ Transformations}\label{LorentzCandP}

Similarly, by handling the variational derivatives ${\mathcal E}_I$ and ${\mathcal E}_{IJ}$ given in \eqref{dervarP1} and \eqref{dervarP2},  we get the off-shell Noether identity 
\begin{equation}
	{\mathcal E}_I \wedge \left ( \tau^I{}_J e^J \right ) + {\mathcal E}_{IJ} \wedge \left ( - D \tau^{IJ} \right ) = d \left [ (-1)^n \tau^{IJ} {\mathcal E}_{IJ} \right ], \label{NILorentz}
\end{equation}
satisfied by infinitesimal local $SO(n-1,1)$ or $SO(n)$ transformations of $e^I$ and $\omega^I{}_J$ with $\tau^{IJ} =- \tau^{JI}$ being the gauge parameter ({\it converse} of Noether's second theorem).

On the other hand, by evaluating the variation \eqref{variation} for an infinitesimal local $SO(n-1,1)$ or $SO(n)$ transformation of $e^I$ and $\omega^I{}_J$, we obtain
\begin{equation}
	\delta_\tau L_P = {\mathcal E}_I \wedge \left ( \tau^I{}_J e^J \right ) + {\mathcal E}_{IJ} \wedge \left ( - D \tau^{IJ} \right ) + d \left [ \kappa \left ( - D \tau^{IJ} \right ) \wedge \star \left ( e_I \wedge e_J \right ) \right ]. \label{variationLorentz}
\end{equation}
Using \eqref{NILorentz}, the right-hand side of \eqref{variationLorentz} acquires the form 
\begin{equation}
	\delta_{\tau} L_P = d \left [ (-1)^n \tau^{IJ} {\mathcal E}_{IJ} + \kappa \left ( - D \tau^{IJ} \right ) \wedge \star \left ( e_I \wedge e_J \right ) \right ]. 
\end{equation}
Once again, the terms inside the brackets can be written as 
\begin{equation}
	(-1)^n \tau^{IJ} {\mathcal E}_{IJ} + \kappa \left ( - D \tau^{IJ} \right ) \wedge \star \left ( e_I \wedge e_J \right ) = d \left [ - \kappa \tau^{IJ} \star \left ( e_I \wedge e_J \right )\right ].
\end{equation}
This off-shell identity has the form
\begin{equation}
	J_{\tau} = d U_{\tau}, \label{keyeq2}
\end{equation}
where we   define  the off-shell current $J_{\tau}$ by 
\begin{equation}
	J_{\tau} := (-1)^n \tau^{IJ} {\mathcal E}_{IJ} + (-1)^{n+1} \kappa \star \left ( e_I \wedge e_J \right ) \wedge \left ( D \tau^{IJ} \right ), \label{JLorentz2}
\end{equation}
and the off-shell Noether potential $U_{\tau}$ as
\begin{equation}
	U_{\tau} := - \kappa \tau^{IJ} \star \left ( e_I \wedge e_J \right ). \label{ULor}
\end{equation}
It is clear from \eqref{keyeq2} that $J_{\tau}$ is off-shell conserved,
\begin{equation}
	d J_{\tau} = 0.
\end{equation} 
It is worth noting that the structure of $J_{\tau}$ in \eqref{JLorentz2} resembles that of the diffeomorphism current \eqref{Jdiff2}.

Therefore, we   show  three things: First, we   apply  our theoretical framework and defined the {\it off-shell} Noether potential $U_{\tau}$ given by \eqref{ULor} and the {\it off-shell} Noether current $J_{\tau}$ given by \eqref{JLorentz2}, both associated to local $SO(n-1,1)$ or $SO(n)$ transformations. Second, we    show  that $U_{\tau}$ and $J_{\tau}$ are related by \eqref{keyeq2}. Third, we    show  that $J_{\tau}$ is {\it off-shell} conserved too.  As in the case of diffeomorphisms, here these three things  are defined {\it off-shell}; we nowhere use the equations of motion in our approach. This differs totally from conventional approaches found in literature (see, e.g.,  \cite{oliveri2020}) which work at the on-shell level only. Thus, we   generalize  and extend these notions to the first-order formalism of general relativity described by the Palatini Lagrangian  and show  that it is {\it not} necessary to define these notions on-shell. 

\subsection{Off-Shell Current for `Improved Diffeomorphisms'}\label{Chafa}
It is pretty obvious that we can combine the relations \eqref{keyeq} and \eqref{keyeq2} involving the off-shell Noether currents and potentials. In particular, by adding them and taking a field-dependent local $SO(n-1,1)$ or $SO(n)$ transformation with gauge parameter $\tau^{IJ}= \zeta \intprod \omega^{IJ}$, we get the off-shell relation 
\begin{equation}
	- \zeta \intprod L_P + (-1)^{n-1} \left ( \zeta \intprod e^I \right ) {\mathcal E}_{I} + \kappa \left ( \zeta \intprod R^{IJ} \right ) \wedge \star \left ( e_I \wedge e_J \right ) =0. \label{NCimproveddiff}
\end{equation}
This off-shell identity is nothing but the current for an `improved diffeomorphism' along a vector field $\zeta$, as we show below. 

In fact, from the variational derivatives ${\mathcal E}_I$ and ${\mathcal E}_{IJ}$ given in \eqref{dervarP1} and \eqref{dervarP2},  we derive the off-shell Noether identity
\begin{equation}
	{\mathcal E}_I \wedge \left [ D \left ( \zeta \intprod e^I \right )+ \zeta \intprod De^I \right ] + {\mathcal E}_{IJ} \wedge \left ( \zeta \intprod R^{IJ} \right ) +d \left [ (-1)^n \left ( \zeta \intprod e^I \right ) {\mathcal E}_{I} \right ]=0, \label{NIID}
\end{equation}
satisfied by the change of $e^I$ and $\omega^I{}_J$ under an `improved diffeomorphism', given by
\begin{eqnarray}
	\delta_{\zeta} e^I & := & D \left ( \zeta \intprod e^I \right )+ \zeta \intprod De^I \nonumber\\
	&= & {\mathcal L}_{\zeta} e^I + \left ( \zeta \intprod \omega^I{}_J \right ) e^J, \\
	\delta_{\zeta} \omega^{IJ} & := & \zeta \intprod R^{IJ} \nonumber\\
	& =& {\mathcal L}_{\zeta} \omega^{IJ} - D \left ( \zeta \intprod \omega^{IJ} \right ),
\end{eqnarray}
which are a linear a combination of a diffeomorphism transformation and a field-dependent local $SO(n-1,1)$ or $SO(n)$ transformation with gauge parameter $\tau^{IJ} = \zeta \intprod \omega^{IJ}$. 

On the other hand, by equating the variation in \eqref{variation} with the change of $e^I$ and $\omega^I{}_J$ under an `improved diffeomorphism', we obtain
\begin{equation}
	\delta_{\zeta} L_P = {\mathcal E}_I \wedge \left [ D \left ( \zeta \intprod e^I \right )+ \zeta \intprod De^I \right ] + {\mathcal E}_{IJ} \wedge \left ( \zeta \intprod R^{IJ} \right ) + d \left [ \kappa \left ( \zeta \intprod R^{IJ} \right ) \wedge \star \left ( e_I \wedge e_J \right ) \right ]\!. \label{variationID}
\end{equation}
Using \eqref{NIID}, the previous expression becomes
\begin{equation}
	\delta_{\zeta} L_P = d \left [ (-1)^{n-1} \left ( \zeta \intprod e^I \right ) {\mathcal E}_{I} + \kappa \left ( \zeta \intprod R^{IJ} \right ) \wedge \star \left ( e_I \wedge e_J \right ) \right ].
\end{equation}
However, the terms inside square brackets can be written as
\begin{equation}
	(-1)^{n-1} \left ( \zeta \intprod e^I \right ) {\mathcal E}_{I} + \kappa \left ( \zeta \intprod R^{IJ} \right ) \wedge \star \left ( e_I \wedge e_J \right )= \zeta \intprod L_P, \label{IID}
\end{equation}
so that for an `improved diffeomorphism' the off-shell Noether current identically vanishes:
\begin{equation}
	J := -\zeta \intprod L_P + (-1)^{n-1} \left ( \zeta \intprod e^I \right ) {\mathcal E}_{I} + \kappa \left ( \zeta \intprod R^{IJ} \right ) \wedge \star \left ( e_I \wedge e_J \right ) =0. \label{NCimproveddiff2}
\end{equation}
Note that \eqref{NCimproveddiff2} is precisely \eqref{NCimproveddiff}.

It is important to remark that the procedure to arrive at \eqref{NCimproveddiff2} differs from the one followed to get the off-shell Noether potential and current for diffeomorphisms presented  in Section \ref{Lied}. The difference relies in the fact that, to arrive at \eqref{NCimproveddiff2},  Cartan's formula is not used at all. If we use it as
\begin{equation}
	{\mathcal L}_{\zeta} \omega^{IJ} = \zeta \intprod R^{IJ} + D \left ( \zeta \intprod \omega^{IJ} \right ), \label{Cartanreescrita}
\end{equation}
then the expression \eqref{IID} is written as
\begin{equation}
	(-1)^{n-1} \left ( \zeta \intprod e^I \right ) {\mathcal E}_{I} + \kappa {\mathcal L}_{\zeta} \omega^{IJ} \wedge \star \left ( e_I \wedge e_J \right ) - 
	\kappa D \left ( \zeta \intprod \omega^{IJ} \right ) \wedge \star \left ( e_I \wedge e_J \right )= \zeta \intprod L_P. \label{missing2}
\end{equation}
Substituting 
\begin{equation}
	\kappa D \left ( \zeta \intprod \omega^{IJ} \right ) \wedge \star \left ( e_I \wedge e_J \right ) = (-1)^n \left ( \zeta \intprod \omega^{IJ} \right ) {\mathcal E}_{IJ} + 
	d \left [ \kappa \left ( \zeta \intprod \omega^{IJ} \right ) \star \left ( e_I \wedge e_J \right ) \right ], 
\end{equation}
into \eqref{missing2} we get precisely \eqref{crucial}, from which \eqref{keyeq} arises. Therefore, from `improved diffeomorphisms' we also obtain the off-shell Noether current and potential associated to diffeomorphisms.

\subsection{Off-Shell Current for the `Generalization of Local Translations'}

It was shown some years ago \cite{Montesinos1} that the variational derivatives ${\mathcal E}_I$ and ${\mathcal E}_{IJ}$ given in~\eqref{dervarP1} and \eqref{dervarP2} can be handled to give the off-shell Noether identity
\begin{equation}
	{\mathcal E}_I \wedge D\rho^I + {\mathcal E}_{IJ} \wedge Z_n{}^{IJ}{}_{KL}\rho^K e^L+d \left [ (-1)^n \rho^I {\mathcal E}_{I} \right ]=0, \label{NILT}
\end{equation}
with
\begin{equation}
	Z_n{}^{IJ}{}_{KL} = {\mathcal R}^{IJ}{}_{KL} - 2\delta^{[I}_K {\mathcal R}^{J]}{}_L +\frac{2}{n-2} \delta^{[I}_L {\mathcal R}^{J]}{}_K + \frac{1}{n-2} \left ( {\mathcal R} + 2 \Lambda \right ) \delta^{[I}_K \delta^{J]}_L.\label{zetan}
\end{equation}
Here, $\rho^I$ is the gauge parameter, ${\mathcal R}^{IJ}{}_{KL}$ are the components of $R^{IJ}$ with respect to the orthonormal frame, $R^{IJ}=(1/2){\mathcal R}^{IJ}{}_{KL} e^K\wedge e^L$; ${\mathcal R}^{I}{}_J:={\mathcal R}^{IK}{}_{JK}$ is the Ricci tensor, and ${\mathcal R}:={\mathcal R}^{I}{}_I$ is the curvature scalar. 

The off-shell Noether identity \eqref{NILT} gives the gauge transformation of $e^I$ and $\omega^I{}_J$ 
\begin{eqnarray}
	\delta_{\rho} e^I &=& D \rho^I, \\
	\delta_{\rho} \omega^{IJ} &=& Z_n{}^{IJ}{}_{KL}\rho^K e^L,
\end{eqnarray}
named `generalization of local translations' because it is the generalization of the so-called `local translations' that exist in 
three dimensions (see \cite{Montesinos2} for a simple derivation of this symmetry in three dimensions and \cite{Montesinos4} for a straightforward derivation in four dimensions).

By computing the variation \eqref{variation} for a `generalization of local translations' of $e^I$ and $\omega^I{}_J$, we have
\begin{equation}
	\delta_{\rho} L_P = {\mathcal E}_I \wedge \delta_{\rho} e^I + {\mathcal E}_{IJ} \wedge \delta_{\rho} \omega^{IJ} + d \left [ \kappa \delta_{\rho} \omega^{IJ} \wedge 
	\star \left ( e_I \wedge e_J \right ) \right ]. \label{variationGLT}
\end{equation}
Using \eqref{NILT}, the right-hand side of last expression becomes
\begin{equation}
	\delta_{\rho} L_P = d\left [ (-1)^{n-1} \rho^I {\mathcal E}_{I}+ \kappa\, Z_n{}^{IJ}{}_{KL}\rho^K e^L \wedge \star \left ( e_I \wedge e_J \right ) \right ].
\end{equation}
If we define the vector field $\rho= \rho^I \partial_I$, where $\partial_I$ is the dual basis of $e^I$ (i.e., $\partial_I \intprod e^J=\delta^J_I$), then $\rho^I=\rho \intprod e^I$. Using this definition, the terms inside square brackets of the previous expression can be expressed as
\begin{eqnarray}
	&&(-1)^{n-1} (\rho \intprod e^I) {\mathcal E}_{I}+ \kappa\, Z_n{}^{IJ}{}_{KL}(\rho \intprod e^K) e^L \wedge \star \left ( e_I \wedge e_J \right ) \nonumber\\
	&&= \rho \intprod L_P+\frac{(-1)^{n-1}}{n-2}(\rho \intprod e^I) {\mathcal E}_{I} - \frac{\sigma}{n-2} \left ( \partial_I \intprod \star {\mathcal E}^I \right ) \rho_J \star e^J. \label{+Palatini}
\end{eqnarray}
Thus, the off-shell Noether current associated to the `generalization of local translations' identically vanishes:
\begin{eqnarray}
	J &:=& (-1)^{n-1} (\rho \intprod e^I) {\mathcal E}_{I} + \kappa\, Z_n{}^{IJ}{}_{KL}(\rho \intprod e^K) e^L \wedge \star \left ( e_I \wedge e_J \right ) \nonumber\\
	&& - \rho\intprod L_P-\frac{(-1)^{n-1}}{n-2}(\rho \intprod e^I) {\mathcal E}_{I} + \frac{\sigma}{n-2} \left ( \partial_I\intprod \star {\mathcal E}^I \right) \rho_J \star e^J \nonumber\\
	&=& 0.\label{JGLT}
\end{eqnarray}
Furthermore, using the off-shell identity
\begin{eqnarray}
	\kappa\, Z_n{}^{IJ}{}_{KL} \left ( \rho \intprod e^K \right ) e^L \wedge \star \left ( e_I \wedge e_J \right ) &=& \left ( \rho\intprod R^{IJ} \right ) \wedge \star \left ( e_I \wedge e_J \right ) + \frac{(-1)^{n-1}}{n-2} \left( \rho \intprod e^I \right ) {\mathcal E}_{I} \nonumber\\
	&& - \frac{\sigma}{n-2} \left ( \partial_I\intprod \star {\mathcal E}^I \right ) \rho_J \star e^J, \label{+++Palatini}
\end{eqnarray} 
the off-shell current \eqref{JGLT} becomes precisely the one given in \eqref{NCimproveddiff2} with $\zeta$ replaced by $\rho$. 

However, note that the identity \eqref{+++Palatini} can be used differently. If \eqref{+++Palatini} is substituted into~\eqref{+Palatini}, we get \eqref{IID} with $\zeta$ replaced by $\rho$, from which \eqref{crucial} and \eqref{keyeq} arise,  as   explained in Section~\ref{Chafa}. Therefore, from the `generalization of local translations',  we also obtain the off-shell Noether current and potential associated to diffeomorphisms.

\section{Holst Lagrangian}\label{Holst}
In four spacetime dimensions, the Holst action \cite{Holst} with a cosmological constant $\Lambda$ is given by the action principle determined by the Lagrangian
\begin{equation}\label{holst}
	L_{H} = \kappa e^I\wedge e^J \wedge \left ( P_{IJKL} R^{KL} - \frac{\Lambda}{12} \epsilon_{IJKL}e^K\wedge e^L\right ),
\end{equation}
where $P_{IJKL}:= (1/2) \epsilon_{IJKL} + (\sigma/\gamma) \eta_{[I|K} \eta_{|J]L}$ and $\gamma\in\mathbb{R}-\{0\}$ is the Immirzi parameter. 

The variation of the Lagrangian (\ref{holst}) under general variations of the independent variables $e^I$ and $\omega^I{}_J$ reads
\begin{equation}\label{varLH}
	\delta L_{H} = {\mathcal E}_I \wedge \delta e^I + {\mathcal E}_{IJ} \wedge \delta \omega^{IJ} + d \left( \kappa P_{IJKL} \delta \omega^{IJ} \wedge e^K\wedge e^L \right),
\end{equation}
where the variational derivatives ${\mathcal E}_I$ and ${\mathcal E}_{IJ}$ are given by
\begin{eqnarray}
	{\mathcal E}_I &:=& - 2 \kappa e^J \wedge \left( P_{IJKL} R^{KL} - \frac{\Lambda}{6} \epsilon_{IJKL}e^K\wedge e^L\right),\label{eqmoth1} \\
	{\mathcal E}_{IJ} &:=& - \kappa D (P_{IJKL} e^K \wedge e^L). \label{eqmoth2}
\end{eqnarray}

\subsection{Off-Shell Current and Potential for Diffeomorphisms}
Using ${\mathcal E}_I$ and ${\mathcal E}_{IJ}$, we obtain the off-shell Noether identity
\begin{eqnarray}\label{NoetherId_diff}
	{\mathcal E}_I \wedge {\mathcal L}_\zeta e^I + {\mathcal E}_{IJ} \wedge {\mathcal L}_\zeta \omega^{IJ} +d \left [ \left ( \zeta\intprod \omega^{IJ} \right ) {\mathcal E}_{IJ} + \left ( \zeta\intprod e^I \right ) {\mathcal E}_I \right ]=0,
\end{eqnarray}
satisfied by the change of $e^I$ and $\omega^{IJ}$ under an infinitesimal diffeomorphism generated by $\zeta$ ({\it converse} of Noether's second theorem).

Then, evaluating the variation (\ref{varLH}) for the change of $e^I$ and $\omega^I{}_J$ under an infinitesimal diffeomorphism generated by $\zeta$, we obtain
\begin{equation}\label{varLH-diff}
	\delta_\zeta L_{H} = {\mathcal E}_I \wedge {\mathcal L}_\zeta e^I + {\mathcal E}_{IJ} \wedge {\mathcal L}_\zeta \omega^{IJ} \! + d \left( \kappa P_{IJKL} {\mathcal L}_\zeta \omega^{IJ} \wedge e^K\wedge e^L \right) \! ,
\end{equation}
which, using (\ref{NoetherId_diff}), is written as
\begin{equation}\label{varLH-diff2}
	\delta_\zeta L_{H} = d\left[ -(\zeta\intprod \omega^{IJ}){\mathcal E}_{IJ} - (\zeta\intprod e^I){\mathcal E}_I \!+ \kappa P_{IJKL} {\mathcal L}_\zeta \omega^{IJ} \wedge e^K\wedge e^L \right] \!.
\end{equation}
Note that the terms inside the brackets can be written as
\begin{eqnarray}
	&&- \left ( \zeta\intprod \omega^{IJ} \right ) {\mathcal E}_{IJ} - \left ( \zeta\intprod e^I \right ) {\mathcal E}_I + \kappa P_{IJKL} {\mathcal L}_\zeta \omega^{IJ} \wedge e^K\wedge e^L \nonumber\\
	&&= \zeta\intprod L_{H} + d\left[ \kappa P_{IJKL} (\zeta\intprod \omega^{IJ}) e^K\wedge e^L \right]. \label{claveHolst+}
\end{eqnarray} 
This off-shell identity has the form
\begin{equation}\label{HolstRel}
	J_{\zeta} = d U_{\zeta},
\end{equation}
where we    define   the off-shell current $J_{\zeta}$ by 
\begin{equation}
	J_{\zeta} := - \zeta\intprod L_{H} -(\zeta\intprod \omega^{IJ}){\mathcal E}_{IJ} - (\zeta\intprod e^I){\mathcal E}_I + \kappa P_{IJKL} {\mathcal L}_\zeta \omega^{IJ} \wedge e^K\wedge e^L , \label{JdiffH}
\end{equation}
and the off-shell Noether potential $U_{\zeta}$ by 
\begin{equation}
	U_{\zeta} :=\kappa P_{IJKL} (\zeta\intprod \omega^{IJ}) e^K\wedge e^L. \label{UdiffH}
\end{equation}
Then, Expression (\ref{HolstRel}) implies that $J_{\zeta}$ is off-shell conserved
\begin{equation}
	d J_{\zeta} = 0.
\end{equation}
Note that the off-shell current \eqref{JdiffH} can be further simplified off-shell, giving
\begin{equation}
	J_{\zeta} = - \left ( \zeta \intprod \omega^{IJ} \right ) {\mathcal E}_{IJ} + \kappa P_{IJKL} \left ( e^I \wedge e^J \right ) 
	\wedge D \left ( \zeta \intprod \omega^{KL} \right ). \label{JdiffH2}
\end{equation}
This expression is relevant because it involves neither ${\mathcal E}_I$ nor $L_H$, in contrast to \eqref{JdiffH}.

In this way, we    show  three things: First, by using our theoretical framework,  we    define   the {\it off-shell} Noether potential $U_{\zeta}$ given by \eqref{UdiffH} and the {\it off-shell} Noether current $J_{\zeta}$ given by \eqref{JdiffH}, which are associated to diffeomorphisms generated by arbitrary vector fields $\zeta$. Second, we    show  that $U_{\zeta}$ and $J_{\zeta}$ are related by \eqref{HolstRel}. Third, we    show  that $J_{\zeta}$ is {\it off-shell} conserved too. To accomplish these three things we   calculate  everything off-shell, and hence our results are different from the existing ones, which are defined only on-shell in literature (see, e.g.,  \cite{Durka_2012,oliveri2020,Freidel1,Freidel2}). Consequently, we   generalize  and extend these notions to the first-order formalism of general relativity described by the Holst Lagrangian  and show  that it is {\it not} necessary to define these notions on-shell. Moreover,  as for the Palatini Lagrangian studied in Section \ref{Lied}, these results coming from the Holst Lagrangian can also be interpreted as a first-order version of the off-shell Noether potential and current associated to diffeomorphisms for general relativity in the metric second-order formalism reported in \cite{Kim1}.  

\subsection{Off-Shell Current and Potential for Local $SO(3,1)$ or $SO(4)$ Transformations}
Similarly, using ${\mathcal E}_I$ and ${\mathcal E}_{IJ}$, we obtain the off-shell Noether identity
\begin{equation}\label{NoetherId_Lorentz}
	{\mathcal E}_I \wedge (\tau^I{}_J e^J) + {\mathcal E}_{IJ} \wedge (-D\tau^{IJ}) = d(\tau^{IJ}{\mathcal E}_{IJ} ),
\end{equation}
satisfied by infinitesimal local $SO(3,1)$ or $SO(4)$ transformations of $e^I$ and $\omega^{IJ}$ with $\tau^{IJ} =- \tau^{JI}$ being the gauge parameter ({\it converse} of Noether's second theorem).

Now, computing the variation (\ref{varLH}) for an infinitesimal local $SO(3,1)$ or $SO(4)$ transformation of $e^I$ and $\omega^I{}_J$, we get
\begin{equation}\label{varLH-Lorentz}
	\delta_\tau L_{H} = {\mathcal E}_I \wedge (\tau^I{}_J e^J) + {\mathcal E}_{IJ} \wedge (-D\tau^{IJ}) + d\left(-\kappa P_{IJKL} D\tau^{IJ}\wedge e^K\wedge e^L \right).
\end{equation}
Using (\ref{NoetherId_Lorentz}), the right-hand side of the previous expression takes the form
\begin{equation}\label{varLH-Lorentz2}
	\delta_\tau L_{H} = d\left(\tau^{IJ}{\mathcal E}_{IJ}-\kappa P_{IJKL} D\tau^{IJ}\wedge e^K\wedge e^L \right).
\end{equation}
The terms inside the parenthesis can be written as
\begin{equation}
	\tau^{IJ}{\mathcal E}_{IJ}-\kappa P_{IJKL} D\tau^{IJ}\wedge e^K\wedge e^L = d\left(-\kappa P_{IJKL} \tau^{IJ} e^K\wedge e^L \right).
\end{equation}
This off-shell identity has the form
\begin{equation}\label{HolstRel2}
	J_{\tau} = d U_{\tau},
\end{equation}
where we    define   the off-shell current $J_{\tau}$ by 
\begin{equation}
	J_{\tau} :=\tau^{IJ}{\mathcal E}_{IJ} - \kappa P_{IJKL} \left ( e^I \wedge e^J \right ) \wedge D \tau^{KL}, \label{JLorH}
\end{equation}
and the off-shell Noether potential $U_{\tau}$ by 
\begin{equation}
	U_{\tau} :=-\kappa P_{IJKL} \tau^{IJ} e^K\wedge e^L. \label{ULorH}
\end{equation}
It follows from (\ref{HolstRel2}) that $J_{\tau}$ is off-shell conserved,
\begin{equation}
	d J_{\tau} = 0.
\end{equation}
Notice that the structure of the $SO(3,1)$ or $SO(4)$ current \eqref{JLorH} resembles that of the diffeomorphism current \eqref{JdiffH2}.

Thus, we    show  three things: First, by applying our theoretical framework,  we    define   the {\it off-shell} Noether potential $U_{\tau}$ given by \eqref{ULorH} and the {\it off-shell} Noether current $J_{\tau}$ given by \eqref{JLorH} associated to local $SO(3,1)$ or $SO(4)$ transformations. Second, we    show  that $U_{\tau}$ and $J_{\tau}$ are related by \eqref{HolstRel2}. Third, we    show  that $J_{\tau}$ is {\it off-shell} conserved too. We remark that, as in the case of diffeomorphisms, these three things are defined off-shell, which is in contrast to the on-shell definitions typically found in literature (see, e.g.,  \cite{oliveri2020}). In this sense, we   generalize  and extend  these notions to the first-order formalism of general relativity described by the Holst Lagrangian  and show  that it is {\it not} necessary to define these notions on-shell. 

\subsection{Off-Shell Current for `Improved Diffeomorphisms'}\label{Chafa2}
By combining the variational derivatives ${\mathcal E}_I$ and ${\mathcal E}_{IJ}$, we obtain the off-shell Noether identity
\begin{equation}\label{NoetherId_imprdiff}
	{\mathcal E}_I \wedge \left [ D \left ( \zeta \intprod e^I \right ) + \zeta \intprod De^I \right ] + {\mathcal E}_{IJ} \wedge \left ( \zeta \intprod R^{IJ} \right ) + d \left[ \left ( \zeta\intprod e^I \right ) {\mathcal E}_I \right] = 0,
\end{equation}
satisfied by the change of $e^I$ and $\omega^I{}_J$ under an `improved diffeomorphism'.

By calculating the variation \eqref{varLH} for the change of $e^I$ and $\omega^I{}_J$ under an `improved diffeomorphism', we have
\begin{eqnarray}\label{varLH_impdiff}
	\delta_\zeta L_{H} &=& {\mathcal E}_I \wedge \left[D(\zeta\intprod e^I) + \zeta\intprod De^I \right] + {\mathcal E}_{IJ} \wedge (\zeta\intprod R^{IJ}) \nonumber\\
	&& + d\left[\kappa P_{IJKL} (\zeta\intprod R^{IJ}) \wedge e^K\wedge e^L \right]. 
\end{eqnarray}
Using (\ref{NoetherId_imprdiff}), the right-hand side of (\ref{varLH_impdiff}) can be written as
\begin{equation}\label{varLH_impdiff2}
	\delta_\zeta L_{H} = d\left[- (\zeta\intprod e^I){\mathcal E}_I + \kappa P_{IJKL} (\zeta\intprod R^{IJ}) \wedge e^K\wedge e^L \right].
\end{equation}
Notice that the terms inside the brackets can be written as
\begin{equation}
	- (\zeta\intprod e^I){\mathcal E}_I + \kappa P_{IJKL} (\zeta\intprod R^{IJ}) \wedge e^K\wedge e^L = \zeta\intprod L_{H}, \label{DiegoG}
\end{equation} 
which means that for an `improved diffeomorphism' the off-shell Noether current identically vanishes:
\begin{equation}
	J := -\zeta\intprod L_{H}-(\zeta\intprod e^I){\mathcal E}_I + \kappa P_{IJKL} (\zeta\intprod R^{IJ}) \wedge e^K\wedge e^L =0. \label{currentHolst_impdiff}
\end{equation}

To close this subsection, we remark that, as for the $n$-dimensional Palatini Lagrangian, we can also use Cartan's identity \eqref{Cartanreescrita} to rewrite \eqref{DiegoG} as
\begin{equation}
	- (\zeta\intprod e^I){\mathcal E}_I + \kappa P_{IJKL} {\mathcal L}_{\zeta} \omega^{IJ} \wedge e^K\wedge e^L - 
	\kappa P_{IJKL} D \left ( \zeta \intprod \omega^{IJ} \right ) \wedge e^K \wedge e^L = \zeta\intprod L_{H}. \label{+Holst}
\end{equation} 
By substituting 
\begin{equation}
	\kappa P_{IJKL} D \left ( \zeta \intprod \omega^{IJ} \right ) \wedge \star \left ( e^K \wedge e^L \right ) = \left ( \zeta \intprod \omega^{IJ} \right ) {\mathcal E}_{IJ} + 
	d \left [ \kappa \left ( \zeta \intprod \omega^{IJ} \right ) \star \left ( e_I \wedge e_J \right ) \right ], 
\end{equation}
into \eqref{+Holst}, we obtain \eqref{claveHolst+}, and then \eqref{HolstRel} arises. Therefore, from `improved diffeomorphisms' we also obtain the off-shell Noether current and potential associated to diffeomorphisms.

\subsection{Off-Shell Current for the `Generalization of Local Translations'}
By handling the variational derivatives ${\mathcal E}_I$ and ${\mathcal E}_{IJ}$, we get the off-shell identity \cite{Montesinos1}
\begin{equation}\label{holstgauge}
	{\mathcal E}_I \wedge D\rho^I + {\mathcal E}_{IJ} \wedge Z{}^{IJ}{}_{KL}\rho^K e^L + d \left ( \rho^I {\mathcal E}_I \right ) =0,
\end{equation}
with
\begin{equation}\label{Zholst}
	Z^{IJ}{}_{KL} = \mathcal{R}^{IJ}{}_{KL} + (P^{-1})^{IJMN} \left( \frac{1}{2} \epsilon_{MNPQ} X^{PQ}{}_{KL} + \frac{\sigma}{\gamma} Y_{MNKL}\right), 
\end{equation}
where we    define  
\begin{eqnarray}
	X_{IJKL}&:=&-2 \eta_{[I|K}{\mathcal R}_{|J]L} + \eta_{[I|L}{\mathcal R}_{|J]K} + \frac12 ( {\mathcal R} + 2 \Lambda ) \eta_{[I|K} \eta_{|J]L},\\
	Y_{IJKL}&:=&\frac12 \left( B_{JLIK} + B_{LIKJ} + B_{IKJL} \right),\label{defX1}
\end{eqnarray}
for $ R^I{}_J \wedge e^J =: (1/3!) B^I{}_{JKL} e^J \wedge e^K \wedge e^L$, and $\rho^I$ is the gauge parameter.

The off-shell Noether identity (\ref{holstgauge}) gives the gauge transformation of $e^I$ and $\omega^{IJ}$
\begin{eqnarray}\label{gaugetrn}
	\delta_{\rho} e^I &&= D \rho^I, \\
	\delta_{\rho} \omega^{IJ} &&= Z{}^{IJ}{}_{KL}\rho^K e^L, 
\end{eqnarray}
named `generalization of local translations.'

By equating the variation in (\ref{varLH}) with a `generalization of local translations' of $e^I$ and $\omega^{IJ}$, we have
\begin{equation}\label{varLH_generalLT}
	\delta_\rho L_{H} = {\mathcal E}_I \wedge \delta_\rho e^I + {\mathcal E}_{IJ} \wedge \delta_{\rho} \omega^{IJ} + d\left(\kappa P_{IJKL} \delta_{\rho} \omega^{IJ} \wedge e^K\wedge e^L \right).
\end{equation}
Then, using (\ref{holstgauge}), the right-hand side of this expression takes the form
\begin{equation}\label{varLH_generalLT2}
	\delta_\rho L_{H} = d\left(- \rho^I {\mathcal E}_I + \kappa P_{IJMN} Z{}^{IJ}{}_{KL}\rho^K e^L \wedge e^M\wedge e^N \right). 
\end{equation}
Defining the vector field $\rho=\rho^I \partial_I$, then $\rho^I=\rho \intprod e^I$ and the terms inside the parenthesis of~\mbox{(\ref{varLH_generalLT2})} can be written as
\begin{eqnarray}
	&&- (\rho \intprod e^I) {\mathcal E}_I + \kappa P_{IJMN} Z{}^{IJ}{}_{KL} (\rho \intprod e^K) e^L \wedge e^M\wedge e^N \nonumber\\
	&&= \rho \intprod L_{H} -\frac{1}{2} (\rho \intprod e^I) {\mathcal E}_I - \frac{\sigma}{2} (\partial_I \intprod \star {\mathcal E}^I) \rho_J \star e^J. \label{+Soren}
\end{eqnarray} 
This implies that for the `generalization of local translations' the off-shell Noether current identically vanishes:
\begin{eqnarray}\label{currentHolst_generalLT}
	J&:=&- (\rho \intprod e^I) {\mathcal E}_I + \kappa P_{IJMN} Z{}^{IJ}{}_{KL} (\rho \intprod e^K) e^L \wedge e^M\wedge e^N \nonumber\\
	&& - \rho \intprod L_{H} + \frac{1}{2} (\rho \intprod e^I) {\mathcal E}_I + \frac{\sigma}{2} (\partial_I \intprod \star {\mathcal E}^I) \rho_J \star e^J \nonumber\\
	& = & 0. 
\end{eqnarray} 
It is worth noting that using the off-shell identity
\begin{eqnarray}
	\kappa P_{IJMN} Z{}^{IJ}{}_{KL} (\rho \intprod e^K) e^L \wedge e^M\wedge e^N &=& \kappa P_{IJKL} (\rho \intprod R^{IJ}) \wedge e^K\wedge e^L \nonumber \\
	&& - \frac{1}{2} (\rho \intprod e^I) {\mathcal E}_I - \frac{\sigma}{2} (\partial_I \intprod \star {\mathcal E}^I) \rho_J \star e^J, \label{+++Soren}
\end{eqnarray}
the off-shell current (\ref{currentHolst_generalLT}) becomes precisely the one given in (\ref{currentHolst_impdiff}) with $\zeta$ replaced by $\rho$.

However, note that the identity \eqref{+++Soren} can be used differently. If \eqref{+++Soren} is substituted into~\eqref{+Soren}, we get \eqref{DiegoG} with $\zeta$ replaced with $\rho$, from which \eqref{claveHolst+} and \eqref{HolstRel} arise,  as   explained in Section \ref{Chafa2}. Therefore, from the `generalization of local translations',  we also obtain the off-shell Noether current and potential associated to diffeomorphisms.

\section{Off-Shell Noether Charges}
An advantage (and a possible use) of the identities satisfied by the off-shell Noether currents and potentials reported in Sections \ref{Palatini} and \ref{Holst} is that they {\it always} hold because no restrictions or specific hypotheses were imposed to obtain them. Therefore, these identities lead naturally to the definition of off-shell Noether charges via
\begin{equation}
	Q = \int_{\Sigma} J = \int_{\partial \Sigma} U, \label{charges}
\end{equation}
where $\Sigma$ is an $(n-1)$-dimensional surface and $\partial \Sigma$ its boundary in the case of the Palatini Lagrangian in $n$-dimensions, whereas $\Sigma$ is a 
three-dimensional surface for the Holst Lagrangian.

It is important to remark that the off-shell Noether charges \eqref{charges} are also {\it kinematical} in the sense that the 
variational derivatives ${\mathcal E}_{IJ}$ and ${\mathcal E}_I$ are not set to zero. Nevertheless, the off-shell currents and potentials are constructed using the $n$-dimensional Palatini and Holst Lagrangians, so they capture or encode the {dynamical} information contained in the Lagrangians through the way the frame $e^I$ and the connection $\omega^I{}_J$ couple to each other. After all, Palatini and Holst Lagrangians lead to the equations of motion for general relativity via the action principle (see \cite{Kim1} for the construction of an off-shell Noether current and potential in the metric second-order formalism). 

The right hand-side of \eqref{charges} can be computed on-shell too, of course, because the off-shell identities between the off-shell potentials and currents are general. Due to the fact that the off-shell potentials for the Holst Lagrangian studied in Section \ref{Holst} depend on the Immirzi parameter, we expect the resulting charges generically depend on this parameter too. Moreover, in Sections \ref{HOS} and \ref{applications},  we   consider the `half off-shell' case for the Holst Lagrangian (in the sense defined there), and we   show  that the Immirzi parameter is present in the resulting expressions for the diffeomorphisms and $SO(3,1)$ or $SO(4)$ potentials and currents. Therefore, from this,  it is deduced that the Immirzi parameter will generically appear in these expressions even `on-shell' too. In fact, the on-shell case is illustrated in Section \ref{applications} and the Immirzi parameter is present. 


\section{Killing Vector Fields}\label{KVF}
If the vector field $\zeta$ is a Killing vector field, then the Lie derivative of the metric tensor along it vanishes,
\begin{equation}
	{\mathcal L}_{\zeta} g=0. \label{Kmetric}
\end{equation}
Since $g=\eta_{IJ} e^I\otimes e^J$, equation \eqref{Kmetric} implies that
\begin{equation}
	\partial^I \intprod {\mathcal L}_{\zeta} e^J =- \partial^J \intprod {\mathcal L}_{\zeta} e^I, 
\end{equation}
which means that the Lie derivative of the orthonormal frame $e^I$ equals an infinitesimal local $SO(n-1,1)$ or $SO(n)$ transformation of itself,
\begin{equation}
	{\mathcal L}_{\zeta} e^I = \tau^I{}_J \left ( \zeta \right ) e^J, \quad \tau^{IJ} \left ( \zeta \right )=-\tau^{JI} \left ( \zeta \right ), \label{Killinge}
\end{equation}
for some suitable gauge parameter $\tau^{IJ} \left ( \zeta \right )$. From this relation,  we obtain, in particular, the field-dependent gauge parameter $\tau^{IJ} \left ( \zeta \right )$
\begin{equation}
	\tau^{IJ} \left ( \zeta \right ) = \partial^J \intprod {\mathcal L}_{\zeta} e^I. \label{taufeo}
\end{equation} 

On the other hand, the Lie derivative of the connection $\omega^I{}_J$ with respect to a Killing vector is more involved because we are working off-shell, and we need to consider separately Palatini and Holst Lagrangians.

\subsection{Palatini Lagrangian}\label{5P}
Using \eqref{dervarP2},  we express $D e^I$ in terms of the variational derivatives ${\mathcal E}_{IJ}$,
\begin{equation}
	D e^I = \frac{\sigma (-1)^{n-1}}{2 \kappa} \bigg[ \star \left ( e^I \wedge {\mathcal E}_{JK} \right ) e^J \wedge e^K + \frac{2}{n-2} \star \left ( e^J \wedge {\mathcal E}_{JK} \right ) e^K \wedge e^I \bigg].
\end{equation}
From this relation and \eqref{Killinge},  we obtain that the Lie derivative of the connection with respect to a Killing vector field equals a local $SO(n)$ or $SO(n-1,1)$ transformation plus a `trivial gauge transformation' $W^{IJ}$ (see \cite{HennBook} for the definition of trivial gauge transformations)
\begin{equation}
	{\mathcal L}_{\zeta} \omega^{IJ} = - D \tau^{IJ} \left ( \zeta \right ) + W^{IJ}, \label{Killingw}
\end{equation} 
where $\tau^{IJ} (\zeta)$ is given by \eqref{taufeo} and
\begin{equation}
	W^{IJ} = {\mathcal L}_{\zeta} K^{IJ} - 2 \tau^{[I}{}_L \left ( \zeta \right ) K^{ \mid L \mid J]}, \quad W^{IJ}=-W^{JI} \label{Wpal}
\end{equation}
with 
\begin{eqnarray}
	K_{IJ} :=& - \frac{\sigma (-1)^{n-1}}{2 \kappa} \bigg[ \star \left ( e_I \wedge {\mathcal E}_{JK} \right ) e^K - \star \left ( e_J \wedge {\mathcal E}_{IK} \right ) e^K \nonumber \\
	& - \star \left ( e_K \wedge {\mathcal E}_{IJ} \right ) e^K - \frac{4}{n-2} \star \left ( e^K \wedge {\mathcal E}_{K [I} \right ) e_{J]} \bigg].\label{Kpal}
\end{eqnarray}
Thus, Expressions \eqref{Killinge} and \eqref{Killingw} are the corresponding changes of the frame $e^I$ and the connection $\omega^I{}_J$ when the vector 
field $\zeta$ is a Killing vector field, for the Palatini Lagrangian~\eqref{PalatiniL}. Equation \eqref{Killingw} expresses the fact that the action of a Killing vector on the connection is compensated by a local $SO(n)$ or $SO(n-1,1)$ transformation and the `trivial gauge transformation' given by the term $W^{IJ}$ proportional to the variational derivative ${\mathcal E}_{IJ}$ according to \eqref{Wpal} and \eqref{Kpal},  thus giving rise to an effective transformation, as shown below.

By substituting \eqref{Killinge} and \eqref{Killingw} into \eqref{NIdiff},  we get the off-shell Noether identity
\begin{eqnarray}
	&&{\mathcal E}_I \wedge \left[ \tau^I{}_J \left ( \zeta \right ) e^J \right ] + {\mathcal E}_{IJ} \wedge \left [ - D \tau^{IJ} \left ( \zeta \right ) + W^{IJ} \right ] \nonumber\\
	&&+ d \left \{ (-1)^n \left [ (\zeta \intprod \omega^{IJ}) {\mathcal E}_{IJ} + (\zeta \intprod e^I ) {\mathcal E}_I \right ] \right \} =0. 
\end{eqnarray}
Using \eqref{NILorentz},  the previous expression acquires the form
\begin{equation}
	{\mathcal E}_{IJ} \wedge W^{IJ} + d \big \{ (-1)^n \big [ \left ( \zeta \intprod \omega^{IJ} + \tau^{IJ} \right ) {\mathcal E}_{IJ} + (\zeta \intprod e^I ) {\mathcal E}_I \big ] \big \} = 0,
\end{equation}
which involves the gauge transformation
\begin{eqnarray}
	\delta e^I &=& 0, \nonumber \\
	\delta \omega^{IJ} &=& W^{IJ}, \label{Ktrivial}
\end{eqnarray}
that leaves the frame $e^I$ unchanged, while the connection $\omega^I{}_J$ undergoes a `trivial gauge transformation'.

By taking the gauge transformation \eqref{Ktrivial} as the starting point and applying the same procedure developed in Section \ref{Palatini}, we get the off-shell relation
\begin{equation}\label{Ohhh}
	J_P = d U_P, 
\end{equation}
with the off-shell potential $U_P$ and the off-shell current $J_P$ defined by 
\begin{eqnarray}
	U_P &:=& \kappa \left [ \tau^{IJ} (\zeta) + \zeta \intprod \omega^{IJ} \right ] \star \left ( e_I \wedge e_J \right ), \\
	J_P &:=& (-1)^{n+1} \left [ \tau^{IJ} (\zeta) + \zeta \intprod \omega^{IJ} \right ] {\mathcal E}_{IJ} + (-1)^n \kappa \star \left ( e_I \wedge e_J \right ) \wedge 
	D \left [ \tau^{IJ} (\zeta) + \zeta \intprod \omega^{IJ} \right ].
\end{eqnarray}
The potential and current can, alternatively, be written as
\begin{eqnarray}
	U_P &=& U_{\zeta} - U_{\tau \left ( \zeta \right )}, \label{Uredu}\\
	J_P &=& J_{\zeta} - J_{\tau \left ( \zeta \right )}, \label{Jredu}
\end{eqnarray} 
where $U_{\zeta}$ is given by \eqref{Udiff} and $J_{\zeta}$ is given by \eqref{Jdiff2}. Similarly, $U_{\tau \left ( \zeta \right )}$ is given by \eqref{ULor} and 
$J_{\tau \left ( \zeta \right )}$ is given by \eqref{JLorentz2} with $\tau^{IJ} \left (\zeta \right )$ given by \eqref{taufeo}. Moreover, it follows from \eqref{Ohhh} that $J_P$ is off-shell conserved,
\begin{equation}
	d J_P =0.
\end{equation}

\subsection{Holst Lagrangian}\label{5H}
Using \eqref{eqmoth2},  we get $D e^I$ in terms of the variational derivatives ${\mathcal E}_{IJ}$
\begin{equation}\label{DeintermsE}
	D e^I = - \frac{\sigma}{2 \kappa} \left [ \star \left ( e^I \wedge E_{JK} \right ) e^J \wedge e^K + \star \left ( e^J \wedge E_{JK} \right ) e^K \wedge e^I \right ],
\end{equation}
with
\begin{equation}
	E_{IJ} := \frac12 \epsilon_{IJKL} \left ( P^{-1} \right )^{KLMN} {\mathcal E}_{MN} = \frac{\gamma^2}{\gamma^2 - \sigma} \left( \delta^K_{[I} \delta^L_{J]} 
	- \frac{1}{2 \gamma} \epsilon_{IJ}{}^{KL} \right ) {\mathcal E}_{KL}.
\end{equation}

From this relation and \eqref{Killinge},  we obtain that the Lie derivative of the connection with respect to a Killing vector field equals a local $SO(3,1)$ or $SO(4)$ transformation plus a `trivial gauge transformation' $W^{IJ} (=- W^{JI})$
\begin{equation}
	{\mathcal L}_{\zeta} \omega^{IJ} = - D \tau^{IJ} \left ( \zeta \right ) + W^{IJ}, \label{KillingwH}
\end{equation} 
where $\tau^{IJ} (\zeta)$ is given by \eqref{taufeo} and
\begin{equation}
	W^{IJ} = {\mathcal L}_{\zeta} K^{IJ} - 2 \tau^{[I}{}_L \left ( \zeta \right ) K^{ \mid L \mid J]}, \label{WHolst}
\end{equation}
with 
\begin{eqnarray}
	K_{IJ} &:= & \frac{\sigma}{2 \kappa} \left[ \star \left ( e_I \wedge E_{JK} \right ) e^K - \star \left ( e_J \wedge E_{IK} \right ) e^K \right. \nonumber \\
	&& \left. - \star \left ( e_K \wedge E_{IJ} \right ) e^K - 2 \star \left ( e^K \wedge E_{K [I} \right ) e_{J]} \right]. \label{KHolst}
\end{eqnarray}
Thus, Expressions \eqref{Killinge} and \eqref{KillingwH} are the corresponding changes of the frame $e^I$ and the connection $\omega^I{}_J$ when the vector 
field $\zeta$ is a Killing vector field, for the Holst Lagrangian~\eqref{holst}. Again, the action of a Killing vector on the connection is compensated by a local $SO(3,1)$ or $SO(4)$ transformation and the `trivial gauge transformation' given by the term $W^{IJ}$ proportional to the variational derivative ${\mathcal E}_{IJ}$ according to \eqref{WHolst} and \eqref{KHolst}, thus, giving rise to an effective transformation, as shown below.

In fact, by substituting \eqref{Killinge} and \eqref{KillingwH} into \eqref{NoetherId_diff},  we get the off-shell Noether identity
\begin{equation}
	{\mathcal E}_I \wedge \left [ \tau^I{}_J \left ( \zeta \right ) e^J \right ] + {\mathcal E}_{IJ} \wedge \left [ - D \tau^{IJ} \left ( \zeta \right ) + W^{IJ} \right ] + d \left [ (\zeta \intprod \omega^{IJ}) {\mathcal E}_{IJ} + (\zeta \intprod e^I ) {\mathcal E}_I \right ] =0. 
\end{equation}
Using \eqref{NoetherId_Lorentz},  the previous expression acquires the form
\begin{equation}
	{\mathcal E}_{IJ} \wedge W^{IJ} + d \left [ \left ( \zeta \intprod \omega^{IJ} + \tau^{IJ} \right ) {\mathcal E}_{IJ} + (\zeta \intprod e^I ) {\mathcal E}_I \right ] = 0,
\end{equation}
which involves the gauge transformation
\begin{eqnarray}
	\delta e^I &=& 0, \nonumber \\
	\delta \omega^{IJ} &=& W^{IJ}, \label{KtrivialH}
\end{eqnarray}
that leaves the frame $e^I$ invariant, while the connection $\omega^I{}_J$ undergoes a `trivial gauge transformation'. 

By taking the gauge transformation \eqref{KtrivialH} as the starting point and applying the same procedure developed in Section \ref{Holst}, we get the off-shell relation
\begin{equation}\label{otra_vez}
	J_H= d U_H, 
\end{equation}
with the off-shell potential $U_H$ and the off-shell current $J_H$ defined by 
\begin{eqnarray}
	U_H &:=& \kappa P_{IJKL} \left [ \tau^{IJ} (\zeta) + \zeta \intprod \omega^{IJ} \right ] \left ( e^K \wedge e^L \right ), \label{UHeffective}\\
	J_H &:=& - \left [ \tau^{IJ} (\zeta) + \zeta \intprod \omega^{IJ} \right ] {\mathcal E}_{IJ} + \kappa P_{IJKL} \left ( e^I \wedge e^J \right ) 
	\wedge D \left [ \tau^{KL} (\zeta) + \zeta \intprod \omega^{KL} \right ]. 
\end{eqnarray}
The potential $U_H $ and current $J_H$ can, alternatively, be written as
\begin{eqnarray}
	U_H &=& U_{\zeta} - U_{\tau \left ( \zeta \right )}, \label{UHredu}\\
	J_H &=& J_{\zeta} - J_{\tau \left ( \zeta \right )}, \label{JHredu}
\end{eqnarray} 
where $U_{\zeta}$ is given by \eqref{UdiffH} and $J_{\zeta}$ is given by \eqref{JdiffH2}. Similarly, $U_{\tau \left ( \zeta \right )}$ is given by \eqref{ULorH} and 
$J_{\tau \left ( \zeta \right )}$ is given by \eqref{JLorH} with $\tau^{IJ} \left (\zeta \right )$ given by \eqref{taufeo}. Furthermore, it follows from \eqref{otra_vez} that $J_H$ is off-shell conserved,
\begin{equation}
	d J_H =0. 
\end{equation}
Note that the Noether potential $U_{\zeta}$ for diffeomorphisms \eqref{UdiffH} when $\zeta$ is a Killing vector field can off-shell, alternatively, be rewritten as
\begin{equation}\label{UHvsUP}
	U_{\zeta} = U^{n=4}_{\zeta} + d \left [ -\frac{\sigma \kappa}{\gamma} \left ( \zeta \intprod e_I \right ) e^I \right ] - \frac{\sigma \kappa}{\gamma} \left ( \tau_{IJ} \left ( \zeta \right ) - \zeta \intprod K_{IJ} \right ) e^I \wedge e^J, 
\end{equation}
where here $U^{n=4}_{\zeta}$ is in fact the corresponding Noether potential \eqref{Udiff} for the Palatini Lagrangian in four-dimensional spacetimes.

Moreover, the potential $U_H$ \eqref{UHeffective} can be further simplified and it acquires the off-shell expression
\begin{equation}
	U_H = U_P + d \left [ - \frac{\sigma \kappa}{\gamma} \left ( \zeta \intprod e_I\right ) e^I \right ] 
	+ \frac{\sigma \kappa}{\gamma} \left \{ \left ( \zeta \intprod e_I \right ) D e^I + \left [ \partial_J \intprod \left ( \zeta \intprod D e_I \right ) \right ] e^I \wedge e^J \right \}, 
\end{equation}
with $U_P$ given by \eqref{Uredu}. Alternatively, using \eqref{DeintermsE}, it can off-shell be written as 
\begin{equation}
	U_H = U_P + d \left [ - \frac{\sigma \kappa}{\gamma} \left ( \zeta \intprod e_I\right ) e^I \right ] + \frac{\sigma\kappa}{\gamma} \left ( \zeta \intprod K_{IJ} \right ) e^I \wedge e^J.
\end{equation}
Therefore,
\begin{equation}
	J_H = d U_H = J_P + d \left [ \frac{\sigma\kappa}{\gamma} \left ( \zeta \intprod K_{IJ} \right ) e^I \wedge e^J \right ],
\end{equation}
with $J_P$ given by \eqref{Jredu}.

\section{Half Off-Shell Case}\label{HOS}
There are essentially three different cases when dealing with gauge symmetries: the first case is defined by ${\mathcal E}_I \neq 0$ and ${\mathcal E}_{IJ} \neq 0 $ and it is named the `off-shell' case. The second case is defined by ${\mathcal E}_I =0$ and ${\mathcal E}_{IJ} = 0$ and it is named the `on-shell' case. The third case is defined by ${\mathcal E}_I \neq 0$ and ${\mathcal E}_{IJ} = 0$, and we name  it the `half off-shell' case (other possible names for this case are `half on-shell' and `semi on-shell'). The `half off-shell' case is the central focus of this section and  is illustrated with examples in Section \ref{applications} to appreciate the explicit expressions for the currents and potentials in spacetimes having particular symmetries. The `on-shell' case is also illustrated in Section \ref{applications}.

In the `half off-shell' case ${\mathcal E}_{IJ}=0$ (and thus  $W^{IJ}=0$). Therefore, $\omega^I{}_J$ becomes the spin connection $\Gamma^I{}_J$, which is defined by 
\begin{equation}
	d e^I + \Gamma^I{}_J \wedge e^J=0, \quad \Gamma_{IJ} =- \Gamma_{JI}, 
\end{equation}
and has the explicit expression
\begin{equation}
	\Gamma^{IJ} = \frac12 \left \{ \partial^J \intprod d e^I - \partial^I \intprod de^J + \left [ \partial^I \intprod \left ( 
	\partial^J \intprod d e^K \right )\right ] e_K \right \}.
\end{equation}

Therefore, both \eqref{Killingw} and \eqref{KillingwH} become
\begin{eqnarray}
	{\mathcal L}_{\zeta} \Gamma^{IJ} &=& - \left ( d \tau^{IJ} + \Gamma^I{}_K \tau^{KJ} + \Gamma^J{}_K \tau^{IK} \right ) \nonumber\\
	& \equiv & - D_{\Gamma} \tau^{IJ}.
\end{eqnarray} 

{\bf Palatini Lagrangian}. In the `half off-shell' case, the expressions for the Noether potentials and currents for the Palatini Lagrangian in $n$-dimensional spacetimes can simply be obtained by replacing ${\mathcal E}_{IJ}=0$ and $\omega^I\,_J$ with $\Gamma^I{}_J$ in the expressions found in Section \ref{Palatini}.

{\bf Holst Lagrangian}. In the `half off-shell' case, the expressions for the Noether potentials and currents for the Holst Lagrangian can simply be obtained by replacing ${\mathcal E}_{IJ}=0$ and $\omega^I\,_J$ with $\Gamma^I{}_J$ in the expressions found in Section \ref{Holst}. Nevertheless, note that the resulting expressions for the potentials and currents for both diffeomorphisms and local $SO(3,1)$ or $SO(4)$ transformations still carry the Immirzi parameter $\gamma$. Therefore, this case is very different from the one we would get if we imposed the `half off-shell'' condition ${\mathcal E}_{IJ}=0$ from the very beginning and we replaced  $\omega^I{}_J$ with $\Gamma^I{}_J$ in the Holst Lagrangian because in such a case the term involving the Immirzi parameter $\gamma$ in $L_H$ would vanish as a consequence of the Bianchi identity, and the Lagrangian $L_H$ would reduce to the Einstein--Hilbert Lagrangian in terms of the frame $e^I$ rather than the metric:
\begin{equation}
	L_{EH} = \kappa \left [ R^{IJ} \wedge \star (e_I \wedge e_J) - 2 \Lambda \eta \right ], 
\end{equation}
with $R^I{}_J = d \Gamma^I{}_J + \Gamma^I{}_K \wedge \Gamma^K{}_J$ being the curvature of $\Gamma^I{}_J$. The action principle defined by this Lagrangian is $S[e]=\int_{\mathcal M} L_{EH}$ and we are in the second-order formalism.

{\bf Einstein--Hilbert Lagrangian}. For the sake of completeness, if we redo the calculations for the Lagrangian $L_{EH}$ for spacetimes in $n$ dimensions, we find the following off-shell Noether identities, potentials, and currents:

{\it Diffeomorphisms}. From the off-shell Noether identity for the change of $e^I$ under an infinitesimal diffeomorphism generated by $\zeta$ 
\begin{equation}
	{\mathcal E}_I \wedge {\mathcal L}_{\zeta} e^I + d \left [ (-1)^n \left ( \zeta \intprod e_I \right ) {\mathcal E}^I \right ] =0, 
\end{equation}
and applying the same off-shell procedure, we get 
\begin{eqnarray}
	U_{\zeta} &:=& \kappa \left ( \zeta \intprod \Gamma^{IJ} \right ) \star \left ( e_I \wedge e_J \right ), \label{UdiffEH} \\
	J_{\zeta} &:=& d U_{\zeta} = (-1)^n \kappa \star \left ( e_I \wedge e_J \right ) \wedge D_{\Gamma} \left ( \zeta \intprod \Gamma^{IJ} \right ). \label{JdiffEH}
\end{eqnarray}

{\it Local $SO(n-1,1)$ or $SO(n)$ transformations}. From the off-shell Noether identity for local $SO(n-1,1)$ or $SO(n)$ transformations 
\begin{equation}
	{\mathcal E}_I \wedge \left ( \tau^I{}_J e^J \right ) =0,
\end{equation}
and applying the same off-shell approach, we obtain
\begin{eqnarray}
	U_{\tau} &:=& - \kappa \tau^{IJ} \star \left ( e_I \wedge e_J \right ), \\
	J_{\tau} &:=& d U_{\tau} = (-1)^{n+1} \kappa \star \left ( e_I \wedge e_J \right ) \wedge D_{\Gamma} \tau^{IJ}. 
\end{eqnarray}
These expressions can alternatively be obtained in an easier way by just setting ${\mathcal E}_{IJ}=0$ (and replacing $\omega^I{}_J$ with $\Gamma^I{}_I$) in the corresponding ones reported in Section \ref{Palatini} of this paper.

{\bf Killing vector fields and half off-shell case}. 
In the `half off-shell' case, we simply have to substitute the corresponding ${\mathcal E}_{IJ}=0$ which implies $K_{IJ}=0$ and $W^{IJ}=0$ in both Sections \ref{5P} and \ref{5H}. Let us analyze more carefully the results for the Holst Lagrangian contained in Section \ref{5H}. In particular, the Noether potential $U_{\zeta}$ for diffeomorphisms \eqref{UHvsUP} when $\zeta$ is a Killing vector field becomes

\begin{equation}
	U_{\zeta} = U^{n=4}_{\zeta} + d \left [ -\frac{\sigma \kappa}{\gamma} \left ( \zeta \intprod e_I \right ) e^I \right ] - \frac{\sigma \kappa}{\gamma} \tau_{IJ} \left ( \zeta \right ) e^I \wedge e^J, \label{UHvsUP2}
\end{equation}
where here $U^{n=4}_{\zeta}$ is in fact the corresponding Noether potential \eqref{Udiff} for the Palatini Lagrangian in four-dimensional spacetimes in the `half off-shell' case too.

Moreover, in `the half off-shell' case, the expressions for the potential $U_H$ and the current $J_P$ for the effective transformation considered in the Section \ref{5H} become 
\begin{eqnarray}
	U_H &=& U_P + d \left [ - \frac{\sigma \kappa}{\gamma} \left ( \zeta \intprod e_I\right ) e^I \right ], \label{UHon} \\
	J_H &=& J_P. \label{JHon}
\end{eqnarray}
We emphasize again that these expressions have been obtained by substituting ${\mathcal E}_{IJ}=0$ at the end of the computations. Notice, however, that the potential $U_H$ for the Holst Lagrangian still carries a $\gamma$ dependence inside the total differential. As   explained  above, this `half off-shell' potential is very different from the potential we would obtain if the condition ${\mathcal E}_{IJ}=0$ were used from the very beginning in the Holst Lagrangian and we replaced $\omega^I{}_J$ with $\Gamma^I{}_J$ in the Holst Lagrangian, because in such a case the term involving the Immirzi parameter would disappear. Notice that,  if the potential \eqref{UHon} is integrated over a two-dimensional compact surface without boundary,   then the last term in \eqref{UHon} vanishes and the charge for the Holst Lagrangian coincides with the charge for the Palatini Lagrangian. 

Finally, we  remark that the potentials \eqref{UHvsUP2} and \eqref{UHon} depend on the Immirzi parameter even in the on-shell case. 

\section{Examples}\label{applications}
In this  section, we apply the theoretical framework developed in Section \ref{HOS} to two relevant spacetimes having particular Killing vector fields, which means that here we restrict our analysis to the `half off-shell' case. It is important to remark that our approach allows us to compute the Noether potentials and currents {\it without} using any specific exact solution of Einstein's equations, only the symmetries of the spacetime are needed. This means that the following expressions {\it cannot} be computed by other approaches, which displays the power of the theoretical framework reported in this paper. 

\subsection{Static Spherically Symmetric Spacetimes}
For the sake of simplicity, let us consider Lorentzian ($\sigma=-1$) spacetimes in four dimensions. The static spherically symmetric spacetime has the metric
\begin{eqnarray}
	g = - f(r) d t \otimes d t + h(r) d r \otimes dr + r^2 \left ( d \theta \otimes d \theta + \sin^2{\theta} \, d \phi \otimes d \phi \right ), \label{SS}
\end{eqnarray}
in local coordinates $x^{\mu}= (x^0,x^1,x^2,x^3) = (t, r, \theta, \phi)$ adapted to the symmetry (static coordinates). 

The orthonormal frame is given by
\begin{eqnarray}
	e^0 = f^{1/2} dt, \quad e^1 = h^{1/2} dr, \quad e^2 = r d \theta, \quad e^3= r \sin{\theta} d \phi. 
\end{eqnarray}
The isometry group of the metric \eqref{SS} is $\mathbb{R}\times SO(3)$ and has associated the following Killing vector fields as generators \cite{BENGURIA197761}:
\begin{eqnarray}
	\zeta_1 &=& \partial_t, \nonumber\\
	\zeta_2 &=& \partial_{\phi}, \nonumber\\
	\zeta_3 &=& \sin{\phi} \, \partial_{\theta} + \cot{\theta} \cos{\phi} \, \partial_{\phi}, \nonumber\\
	\zeta_4 &=& - \cos{\phi} \, \partial_{\theta} + \cot{\theta} \sin{\phi} \, \partial_{\phi}.
\end{eqnarray}
The vector $\zeta_1$ is the generator of time translations, whereas the vectors $\zeta_2$, $\zeta_3$, and $\zeta_4$ correspond to the components of the angular momentum, that is, to the generators of $SO(3)$. From now on, we take $f(r) = e^{2 a (r)}$ and $h(r)= e^{2 b(r)}$ to simplify the calculations. With this at hand, the `half off-shell' potentials and currents for diffeomorphisms associated with the Killing vector fields, local $SO(3,1)$ transformations induced by Killing vector fields, and the effective transformation acquire the explicit forms contained in \linebreak Sections \ref{PalSph} and \ref{HolstDLE}.

\subsubsection{Palatini Lagrangian}\label{PalSph}

\subsubsection*{ Half Off-Shell Potentials and Currents for Diffeomorphisms Generated by the Killing Vector Fields} 
\begin{enumerate}
	\item[(i)] For the Killing vector field $\zeta_1$, the potential \eqref{Udiff} acquires the form
	\begin{equation}
		U_{\zeta_1} = 2 \kappa \left (\frac{d a}{d r} \right ) \, e^{ \left ( a-b \right )}e^2 \wedge e^3, 
	\end{equation}
	and therefore
	\begin{eqnarray}
		J_{\zeta_1} & = & d U_{\zeta_1} \nonumber\\
		& = & 2 \kappa \frac{e^{\left ( a- 2 b \right ) }}{r} \bigg[ r \left ( \frac{d a}{dr}\right )^2 + \frac{d a}{d r} \left (2 - r \frac{d b}{d r} \right ) + r \frac{d^2 a}{d r^2} \bigg] e^1 \wedge e^2 \wedge e^3.
	\end{eqnarray}
	\item[(ii)] Likewise, for the Killing vector field $\zeta_2$, the potential \eqref{Udiff} becomes
	\begin{eqnarray}
		U_{\zeta_2} = - 2 \kappa \cos{\theta} \, e^0 \wedge e^1 + 2 \kappa e^{-b} \sin{\theta} \, e^0 \wedge e^2, 
	\end{eqnarray}
	and so
	\begin{eqnarray}
		J_{\zeta_2} & = & d U_{\zeta_2} \nonumber\\
		& = & 2 \kappa \frac{e^{- 2 b }}{r} \sin{\theta} \bigg[ -1 + e^{2b} + r \left ( \frac{d b}{dr} - \frac{d a}{d r} \right ) \bigg] e^0 \wedge e^1 \wedge e^2.
	\end{eqnarray}
	\item[(iii)] For the Killing vector field $\zeta_3$, the potential \eqref{Udiff} acquires the form
	\begin{eqnarray}
		U_{\zeta_3} &=& - 2 \kappa \cos{\theta} \cot{\theta} \cos{\phi} \, e^0 \wedge e^1 + 2 \kappa e^{- b} \cos{\theta} \cos{\phi} \, e^0 \wedge e^2 \nonumber\\
		&& - 2 \kappa e^{-b} \sin{\phi} \, e^0 \wedge e^3,
	\end{eqnarray}
	and then
	\begingroup\makeatletter\def\f@size{9}\check@mathfonts
	\def\maketag@@@#1{\hbox{\m@th\normalsize\normalfont#1}}%
	\begin{eqnarray}
		J_{\zeta_3} &= & d U_{\zeta_3} \nonumber\\
		& = & 2 \kappa \frac{e^{-2b}}{r} \cos{\phi} \bigg \{ e^{2b} \csc{\theta} \cot{\theta} - \left [ 1 - e^{2 b} + r \left( \frac{d a}{d r} - \frac{d b}{d r} \right ) \right ]\cos{\theta} \bigg \} e^0 \wedge e^1 \wedge e^2 \nonumber\\
		& & + 2 \kappa \frac{e^{-2b}}{r} \sin{\phi} \bigg [ 1 + e^{2b} \cot^2{\theta} + r \left ( \frac{d a}{d r} - \frac{d b}{d r} \right ) \bigg ] e^0 \wedge e^1 \wedge e^3.
	\end{eqnarray}
	\endgroup
	\item[(iv)] For the Killing vector field $\zeta_4$, the potential \eqref{Udiff} becomes
	\begin{eqnarray}
		U_{\zeta_4} &=& - 2 \kappa \cos{\theta} \cot{\theta} \sin{\phi} \, e^0 \wedge e^1 + 2 \kappa e^{- b} \cos{\theta} \sin{\phi} \, e^0 \wedge e^2 \nonumber\\
		&& + 2 \kappa e^{-b} \cos{\phi} \, e^0 \wedge e^3,
	\end{eqnarray}
	and therefore
	\begingroup\makeatletter\def\f@size{9}\check@mathfonts
	\def\maketag@@@#1{\hbox{\m@th\normalsize\normalfont#1}}%
	\begin{eqnarray}
		J_{\zeta_4} &=& d U_{\zeta_4} \nonumber\\
		& = & 2 \kappa \frac{e^{-2b}}{r} \sin{\phi} \bigg \{ - \left [ 1 - e^{2b} + r \left ( \frac{d a}{d r} - \frac{d b}{d r} \right ) \right ] \cos{\theta} + e^{2 b} \cot{\theta} \csc{\theta} \bigg \} e^0 \wedge e^1 \wedge e^2 \nonumber \\
		& & - 2 \kappa \frac{e^{- 2 b}}{r} \cos{\phi} \bigg [ 1 + e^{2 b} \cot^2{\theta} + r \left ( \frac{d a}{d r} - \frac{d b}{d r} \right ) \bigg ] e^0 \wedge e^1 \wedge e^3.
	\end{eqnarray}
	\endgroup
\end{enumerate}

We can do more. We can compute the integral of the potential $U_{\zeta_1}$ over a sphere $S^2$ of constant radius $r$, which defines the conserved charge inside it. We obtain the `half off-shell' charge
\begin{equation}
	Q (r) = \int_{S^2} U_{\zeta_1} = 8 \pi \kappa r^2 e^{ \left [ a(r) - b(r) \right ]} \frac{da}{dr}.
\end{equation}
This result is general. For instance, using in particular the explicit expressions 
\begin{eqnarray}
	e^{2 a} &=& 1 -\frac{2 M}{r} - \frac{\Lambda r^2}{3}, \nonumber\\
	e^{- 2 b} &=& 1 -\frac{2 M}{r} - \frac{\Lambda r^2}{3}, \label{ab}
\end{eqnarray}
with $M$ the ``mass parameter'' and $\Lambda$ the cosmological constant, which correspond to the Schwarzschild--de-Sitter or the Schwarzschild--anti-de-Sitter solution depending on the sign of $\Lambda$, we arrive at the on-shell charge
\begin{equation}
	Q (r) = \int_S U_{\zeta_1} = 8 \pi \kappa \left ( M - \frac{\Lambda r^3}{3} \right ).
\end{equation}
Note that the term involving $\Lambda$ has a very appealing behavior. Such a term is added to $M$ if $\Lambda<0$ while it is subtracted from $M$ if $\Lambda >0$, thus indicating an attractive effect in the former case (effective mass increases) and a repulsive effect in the latter case (effective mass decreases). Of course, the region of spacetime and its boundary must be clearly defined to calculate the Noether charges using the potentials computed in this paper, and the current calculations can be used to achieve that goal. In particular, it would be interesting to use the current expressions to compute masses, energies, and entropy of the Schwarzschild--de-Sitter black hole, and compare with the results of   \citet{Corichi1}. Similarly, the Schwarzschild--anti-de-Sitter black hole can be analyzed and compared with the results in  \cite{PhysRevLett.84.1647,Durka_2012}.

\subsubsection*{ Half Off-Shell Potentials and Currents for Local $SO(3,1)$ Transformations Induced by Killing Vector Fields} 
(i) For the gauge parameter $\tau^{IJ} \left ( \zeta_1 \right )$, the potential \eqref{ULor} and its current acquire the form
\begin{eqnarray}
	U_{\tau \left ( \zeta_1 \right )} = 0, \quad J_{\tau \left ( \zeta_1 \right )} = 0.
\end{eqnarray}
(ii) For the gauge parameter $\tau^{IJ} \left ( \zeta_2 \right )$, the potential \eqref{ULor} and its current become
\begin{eqnarray}
	U_{\tau \left ( \zeta_2 \right )} = 0, \quad J_{\tau \left ( \zeta_2 \right )} = 0.
\end{eqnarray}
This is so because both $\tau^{IJ} \left ( \zeta_1 \right )$ and $\tau^{IJ} \left ( \zeta_2 \right )$ vanish.\\
(iii) For the gauge parameter $\tau^{IJ} \left ( \zeta_3 \right )$, the potential \eqref{ULor} acquires the form
\begin{equation}
	U_{\tau \left ( \zeta_3 \right )} = - 2 \kappa \csc{\theta} \cos{\phi} \, e^0 \wedge e^1,
\end{equation}
and thus
\begin{eqnarray}
	J_{\tau \left ( \zeta_3 \right )} &=& d U_{\tau \left ( \zeta_3 \right )}\nonumber\\
	& = & \frac{2 \kappa}{r} \cot{\theta} \csc{\theta} \cos{\phi} \, e^0 \wedge e^1 \wedge e^2 + \frac{2 \kappa}{r} {\csc}^2 {\theta} \sin{\phi} \, e^0 \wedge e^1 \wedge e^3. 
\end{eqnarray}
(iv) For the gauge parameter $\tau^{IJ} \left ( \zeta_4 \right )$, the potential \eqref{ULor} becomes
\begin{eqnarray}
	U_{\tau \left ( \zeta_4 \right )} = - 2 \kappa \csc{\theta} \sin{\phi} \, e^0 \wedge e^1,
\end{eqnarray}
and thus
\begin{eqnarray}
	J_{\tau \left ( \zeta_4 \right )} &=& d U_{\tau \left ( \zeta_4 \right )} \nonumber\\
	& = & \frac{2 \kappa}{r} \cot{\theta} \csc{\theta} \sin{\phi} \, e^0 \wedge e^1 \wedge e^2 - \frac{2 \kappa}{r} {\csc}^2 {\theta} \cos{\phi} \, e^0 \wedge e^1 \wedge e^3. 
\end{eqnarray}

\subsubsection{Holst Lagrangian}\label{HolstDLE}

\subsubsection*{ Half Off-Shell Potentials and Currents for Diffeomorphisms Generated by the Killing Vector Fields} 
(i) For the Killing vector field $\zeta_1$, the potential \eqref{UHvsUP2} acquires the form
\begin{equation}
	U_{\zeta_1} = 2 \kappa \left (\frac{d a}{d r} \right ) \, e^{ \left ( a-b \right )}e^2 \wedge e^3 + \frac{2 \kappa}{\gamma} 
	\left (\frac{d a}{d r} \right ) \, e^{ \left ( a-b \right )} e^0 \wedge e^1, 
\end{equation}
and thus
\begin{eqnarray}
	J_{\zeta_1} & = & d U_{\zeta_1} \nonumber\\
	& = & 2 \kappa \frac{e^{\left ( a- 2 b \right ) }}{r} \bigg[ r \left ( \frac{d a}{dr}\right )^2 + \frac{d a}{d r} \left (2 - r \frac{d b}{d r} \right ) + r \frac{d^2 a}{d r^2} \bigg] e^1 \wedge e^2 \wedge e^3.
\end{eqnarray}
Note that there is no $\gamma$ in the current $J_{\zeta_1}$ despite the fact that it appears in the potential $U_{\zeta_1}$.\\
(ii) For the Killing vector field $\zeta_2$, the potential \eqref{UHvsUP2} becomes
\begin{eqnarray}
	U_{\zeta_2} &=& - 2 \kappa \cos{\theta} \, e^0 \wedge e^1 + 2 \kappa e^{-b} \sin{\theta} \, e^0 \wedge e^2 \nonumber\\
	&& + \frac{2 \kappa}{\gamma} \cos{\theta}\, e^2 \wedge e^3 + \frac{2 \kappa}{\gamma} e^{-b} \sin{\theta}\, e^1 \wedge e^3,
\end{eqnarray}
and so
\begin{eqnarray}
	J_{\zeta_2} & = & d U_{\zeta_2} \nonumber\\
	& = & 2 \kappa \frac{e^{- 2b }}{r} \sin{\theta} \bigg[ -1 + e^{2b} + r \left ( \frac{d b}{dr} - \frac{d a}{d r} \right ) \bigg] e^0 \wedge e^1 \wedge e^2.
\end{eqnarray}
Note that there is no $\gamma$ in the current $J_{\zeta_2}$  even though  it is in the potential $U_{\zeta_2}$.\\
(iii) For the Killing vector field $\zeta_3$, the potential \eqref{UHvsUP2} acquires the form 
\begingroup\makeatletter\def\f@size{9}\check@mathfonts
\def\maketag@@@#1{\hbox{\m@th\normalsize\normalfont#1}}%
\begin{eqnarray}
	U_{\zeta_3} &=& - 2 \kappa \cos{\theta} \cot{\theta} \cos{\phi} \, e^0 \wedge e^1 + 2 \kappa e^{- b} \cos{\theta} \cos{\phi} \, e^0 \wedge e^2 - 2 \kappa e^{-b} \sin{\phi} \, e^0 \wedge e^3 \nonumber\\
	& &+ \frac{2 \kappa}{\gamma} \cos{\theta} \cot{\theta} \cos{\phi} \, e^2 \wedge e^3 + \frac{2 \kappa}{\gamma} e^{- b} \cos{\theta} \cos{\phi} \, e^1 \wedge e^3 + \frac{2 \kappa}{\gamma} e^{-b} \sin{\phi} \, e^1 \wedge e^2,\nonumber\\
	& &
\end{eqnarray}
\endgroup
and then
\begin{eqnarray}
	J_{\zeta_3} &= & d U_{\zeta_3} \nonumber\\
	& = & 2 \kappa \frac{e^{-2b}}{r} \cos{\phi} \bigg \{ e^{2b} \csc{\theta} \cot{\theta} - \left [ 1 - e^{2 b} + r \left ( \frac{d a}{d r} - \frac{d b}{d r} \right ) \right ] \cos{\theta} \bigg \} e^0 \wedge e^1 \wedge e^2 \nonumber\\
	& & + 2 \kappa \frac{e^{-2b}}{r} \sin{\phi} \bigg [ 1 + e^{2b} \cot^2{\theta} + r \left ( \frac{d a}{d r} - \frac{d b}{d r} \right ) \bigg ] e^0 \wedge e^1 \wedge e^3 \nonumber\\
	& & + \frac{4 \kappa}{\gamma r} e^{- b} \csc{\theta} \cos{\phi} \, e^1 \wedge e^2 \wedge e^3. 
\end{eqnarray}
(iv) For Killing vector field $\zeta_4$, the potential \eqref{UHvsUP2} becomes\begingroup\makeatletter\def\f@size{9}\check@mathfonts
\def\maketag@@@#1{\hbox{\m@th\normalsize\normalfont#1}}%
\begin{eqnarray}
	U_{\zeta_4} &=& - 2 \kappa \cos{\theta} \cot{\theta} \sin{\phi} \, e^0 \wedge e^1 + 2 \kappa e^{- b} \cos{\theta} \sin{\phi} \, e^0 \wedge e^2 + 2 \kappa e^{-b} \cos{\phi} \, e^0 \wedge e^3 \nonumber\\
	& & + \frac{2 \kappa}{\gamma} \cos{\theta} \cot{\theta} \sin{\phi} \, e^2 \wedge e^3 + \frac{2 \kappa}{\gamma} e^{- b} \cos{\theta} \sin{\phi} \, e^1 \wedge e^3 - \frac{2 \kappa}{\gamma} e^{-b} \cos{\phi} \, e^1 \wedge e^2,\nonumber\\
	& &
\end{eqnarray}
\endgroup
and therefore
\begin{eqnarray}
	J_{\zeta_4} &=& d U_{\zeta_4} \nonumber\\
	& = & 2 \kappa \frac{e^{-2b}}{r} \sin{\phi} \bigg \{ - \left [ 1 - e^{2b} + r \left ( \frac{d a}{d r} - \frac{d b}{d r} \right ) \right
	] \cos{\theta} + e^{2 b} \cot{\theta} \csc{\theta} \bigg \} e^0 \wedge e^1 \wedge e^2 \nonumber \\
	& & - 2 \kappa \frac{e^{- 2 b}}{r} \cos{\phi} \bigg [ 1 + e^{2 b} \cot^2{\theta} + r \left ( \frac{d a}{d r} - \frac{d b}{d r} \right ) \bigg ] e^0 \wedge e^1 \wedge e^3 
	\nonumber\\
	&& + \frac{4 \kappa}{\gamma r} e^{-b} \csc{\theta} \sin{\phi} e^1 \wedge e^2 \wedge e^3.
\end{eqnarray}

Notice that in  Cases (iii) and (iv) the Immirzi parameter shows up in the corresponding expressions for potentials and currents. Even if the explicit expressions \eqref{ab} are used, the Immirzi parameter will be present.

\subsubsection*{ Half Off-Shell Potentials and Currents for Local $SO(3,1)$ Transformations Induced by the Killing Vector Fields} 
\begin{enumerate}
	
	\item[(i)] For the gauge parameter $\tau^{IJ} \left ( \zeta_1 \right )$, the potential \eqref{ULorH} and its current acquire the form
	\begin{eqnarray}
		U_{\tau \left ( \zeta_1 \right )}= 0 , \quad J_{\tau \left ( \zeta_1 \right )}=0. 
	\end{eqnarray}
	\item[(ii)] For the gauge parameter $\tau^{IJ} \left ( \zeta_2 \right )$, the potential \eqref{ULorH} and its current become
	\begin{eqnarray}
		U_{\tau \left ( \zeta_2 \right )}= 0 , \quad J_{\tau \left ( \zeta_2 \right )}=0. 
	\end{eqnarray}
	This is so because both $\tau^{IJ} \left ( \zeta_1 \right )$ and $\tau^{IJ} \left ( \zeta_2 \right )$ vanish.\\
	\item[(iii)] For the gauge parameter $\tau^{IJ} \left ( \zeta_3 \right )$, the potential \eqref{ULorH} acquires the form
	\begin{equation}
		U_{\tau \left ( \zeta_3 \right )} = - 2 \kappa \csc{\theta} \cos{\phi} \, e^0 \wedge e^1 + \frac{2 \kappa}{\gamma} \csc{\theta} \cos{\phi} \, e^2 \wedge e^3,
	\end{equation}
	and thus
	\begin{eqnarray}
		J_{\tau \left ( \zeta_3 \right )} &=& d U_{\tau \left ( \zeta_3 \right )}\nonumber\\
		& = & \frac{2 \kappa}{r} \cot{\theta} \csc{\theta} \cos{\phi} \, e^0 \wedge e^1 \wedge e^2 + \frac{2 \kappa}{r} {\csc}^2 {\theta} \sin{\phi} \, e^0 \wedge e^1 \wedge e^3 \nonumber\\
		& & + \frac{4 \kappa}{\gamma r} e^{-b} \csc{\theta} \cos{\phi} \, e^1 \wedge e^2 \wedge e^3.
	\end{eqnarray}
	\item[(iv)] For the gauge parameter $\tau^{IJ} \left ( \zeta_4 \right )$, the potential \eqref{ULorH} becomes
	\begin{equation}
		U_{\tau \left ( \zeta_4 \right )} = - 2 \kappa \csc{\theta} \sin{\phi} \, e^0 \wedge e^1 + \frac{2 \kappa}{\gamma} \csc{\theta} \sin{\phi} \, e^2 \wedge e^3,
	\end{equation}
	and thus
	\begin{eqnarray}
		J_{\tau \left ( \zeta_4 \right )} &=& d U_{\tau \left ( \zeta_4 \right )} \nonumber\\
		& = & \frac{2 \kappa}{r} \cot{\theta} \csc{\theta} \sin{\phi} \, e^0 \wedge e^1 \wedge e^2 - \frac{2 \kappa}{r} {\csc}^2 {\theta} \cos{\phi} \, e^0 \wedge e^1 \wedge e^3 \nonumber\\
		& & + \frac{4\kappa}{\gamma r} e^{-b} \csc{\theta} \sin{\phi} e^1 \wedge e^2 \wedge e^3.
	\end{eqnarray}
\end{enumerate}

Again, the Immirzi parameter shows up in the potentials and in their associated currents, and it is not possible to get rid of it even in the `on-shell' case. 

\subsubsection*{ Half Off-Shell Potentials and Currents \eqref{UHon} and \eqref{JHon}}
\begin{enumerate}
	\item[(i)] For $\zeta_1$
	\begin{eqnarray}
		U_H &=& U_{\zeta_1} \nonumber\\
		&=&U_P+d\left[ \frac{\kappa }{\gamma} (\zeta_1 \intprod e_I) e^I \right] \nonumber\\
		&=& 2 \kappa \left (\frac{d a}{d r} \right ) \, e^{ \left ( a-b \right )}e^2 \wedge e^3 + d\left(-\frac{ \kappa }{\gamma } e^a \, e^0 \right), \label{SS-UH-1}\\
		J_H &=& d U_H = J_P = J_{\zeta_1} \nonumber\\
		& = & 2 \kappa \frac{e^{\left ( a- 2 b \right ) }}{r} \bigg[ r \left ( \frac{d a}{dr}\right )^2 + \frac{d a}{d r} \left (2 - r \frac{d b}{d r} \right ) + r \frac{d^2 a}{d r^2} \bigg] e^1 \wedge e^2 \wedge e^3. 
	\end{eqnarray}
	Notice that $U_P$ is given by the first term in the last equality in \eqref{SS-UH-1}.\\
	\item[(ii)] For $\zeta_2$
	\begin{eqnarray}
		U_H &=& U_{\zeta_2} \nonumber\\
		&=&U_P+d\left[ \frac{\kappa }{\gamma} (\zeta_2 \intprod e_I) e^I \right] \nonumber\\
		&=& - 2 \kappa \cos{\theta} \, e^0 \wedge e^1 + 2 \kappa e^{-b} \sin{\theta} \, e^0 \wedge e^2 + d\left( \frac{\kappa}{\gamma } r \sin{\theta} \, e^3 \right), \label{SS-UH-2}\\
		J_H &=& d U_H = J_P = J_{\zeta_2}\nonumber\\
		& = & 2 \kappa \frac{e^{- 2 b }}{r} \sin{\theta} \bigg[ -1 + e^{2b} + r \left ( \frac{d b}{dr} - \frac{d a}{d r} \right ) \bigg] e^0 \wedge e^1 \wedge e^2.
	\end{eqnarray}
	Notice that $U_P$ is given by the first and second terms in the last equality in \eqref{SS-UH-2}.\\
	\item[(iii)] For $\zeta_3$
	\begin{eqnarray}
		U_H &=& U_{\zeta_3}- U_{\tau \left ( \zeta_3 \right )} \nonumber\\
		&=&U_P+d\left[ \frac{\kappa }{\gamma} (\zeta_3 \intprod e_I) e^I \right] \nonumber\\
		&=& 2 \kappa \sin{\theta} \cos{\phi} \, e^0 \wedge e^1 + 2 \kappa e^{- b} \cos{\theta} \cos{\phi} \, e^0 \wedge e^2 - 2 \kappa e^{-b} \sin{\phi} \, e^0 \wedge e^3 \nonumber\\
		&&+d\left( \frac{\kappa r}{\gamma } \sin\phi \, e^2 + \frac{\kappa r}{\gamma } \cos\theta \cos\phi \, e^3\right), \label{SS-UH-3} \\
		J_H &=& d U_H = J_P = J_{\zeta_3}- J_{\tau \left ( \zeta_3 \right )}\nonumber \\
		& = & 2 \kappa \frac{e^{-2b}}{r} \cos{\theta} \cos{\phi} \bigg ( -1 + e^{2b} - r \frac{d a}{d r} + r \frac{d b}{d r} \bigg ) e^0 \wedge e^1 \wedge e^2 \nonumber\\
		& & - 2 \kappa \frac{e^{-2b}}{r} \sin{\phi} \bigg ( -1 + e^{2b} - r \frac{d a}{d r} + r \frac{d b}{d r} \bigg ) \, e^0 \wedge e^1 \wedge e^3.
	\end{eqnarray}
	Notice that $U_P$ is given by the first three terms in the last equality in \eqref{SS-UH-3}.\\
	\item[(iv)] For $\zeta_4$
	\begin{eqnarray}
		U_H &=& U_{\zeta_4}- U_{\tau \left ( \zeta_4 \right )} \nonumber\\
		&=&U_P+d\left[ \frac{\kappa }{\gamma} (\zeta_4 \intprod e_I) e^I \right] \nonumber\\
		&=& 2 \kappa \sin{\theta} \sin{\phi} \, e^0 \wedge e^1 + 2 \kappa e^{- b} \cos{\theta} \sin{\phi} \, e^0 \wedge e^2 + 2 \kappa e^{-b} \cos{\phi} \, e^0 \wedge e^3 \nonumber\\
		&&+d\left( -\frac{\kappa r}{\gamma } \cos\phi \, e^2 + \frac{\kappa r}{\gamma } \cos\theta \sin\phi \, e^3\right), \label{SS-UH-4}\\
		J_H &=& d U_H = J_P = J_{\zeta_4}- J_{\tau \left ( \zeta_4 \right )}\nonumber \\
		& = & 2 \kappa \frac{e^{-2b}}{r} \cos{\theta} \sin{\phi} \bigg ( -1 + e^{2b} - r \frac{d a}{d r} + r \frac{d b}{d r} \bigg ) e^0 \wedge e^1 \wedge e^2 \nonumber\\
		& & + 2 \kappa \frac{e^{-2b}}{r} \cos{\phi} \bigg ( -1 + e^{2b} - r \frac{d a}{d r} + r \frac{d b}{d r} \bigg ) \, e^0 \wedge e^1 \wedge e^3.
	\end{eqnarray}
	Notice that $U_P$ is given by the first three terms in the last equality in \eqref{SS-UH-4}.
\end{enumerate}

\subsection{Friedmann--Lemaitre--Robertson--Walker Cosmology}

For the sake of simplicity, let us consider Lorentzian ($\sigma=-1$) spacetimes in four dimensions with homogeneous and isotropic spacelike slices. In local coordinates \linebreak \mbox{$x^{\mu}= (x^0,x^1,x^2,x^3) = (t, r, \theta, \phi)$} adapted to these symmetries, the general form of the metric is given by the FLRW metric
\begin{equation}
	g= - d t \otimes d t + a^2 (t) \left ( \frac{1}{1- k r^2} d r \otimes d r + r^2 d \theta \otimes d \theta + r^2 \sin^2{\theta} \, d \phi \otimes d \phi \right ), \label{FLRWmetric}
\end{equation}
where $k =0,1,-1$ is the spatial curvature and $a(t)$ is the scale factor. Do not confuse $k$ with $\kappa$ in the expressions of this Subsection.

From \eqref{FLRWmetric},  we read off the orthonormal frame given by
\begin{equation}
	e^0 = dt, \quad e^1 = \frac{a (t)}{\sqrt{1 - k r^2}} dr, \quad e^2 = a(t) r d \theta, \quad e^3= a(t) r \sin{\theta} d \phi. 
\end{equation}
Since the spatial part of \eqref{FLRWmetric} is maximally symmetric, it has associated the following six Killing vector fields \cite{Jambrina}
\begin{eqnarray}
	\chi_1 &=& \sqrt{1 - k r^2} \left ( \sin{\theta} \cos{\phi} \frac{\partial}{\partial r} + \frac{1}{r} \cos{\theta} \cos{\phi} 
	\frac{\partial}{\partial \theta} - \frac{1}{r} \csc{\theta} \sin{\phi} \frac{\partial}{\partial \phi} \right ), \nonumber\\
	\chi_2 &=& \sqrt{1 - k r^2} \left ( \sin{\theta} \sin{\phi} \frac{\partial}{\partial r} + \frac{1}{r} \cos{\theta} \sin{\phi} 
	\frac{\partial}{\partial \theta} + \frac{1}{r} \csc{\theta} \cos {\phi} \frac{\partial}{\partial \phi} \right ), \nonumber\\
	\chi_3 &=& \sqrt{1 - k r^2} \left ( \cos{\theta} \frac{\partial}{\partial r} - \frac{1}{r} \sin{\theta} \frac{\partial}{\partial \theta} \right ), \nonumber\\
	\zeta_1 &=& \sin{\phi} \frac{\partial}{\partial \theta} + \cot{\theta} \cos{\phi} \frac{\partial}{\partial \phi}, \nonumber\\
	\zeta_2 &=& \cos{\phi} \frac{\partial}{\partial \theta} - \cot{\theta} \sin{\phi} \frac{\partial}{\partial \phi}, \nonumber\\
	\zeta_3 &=& \frac{\partial}{\partial \phi}.
\end{eqnarray}
The Lie algebra of the Killing vector fields ($\zeta_{AB}=- \zeta_{BA}$) is the following:
\begin{equation}
	\left [ \zeta_{AB}, \zeta_{CD} \right ] = g_{AC} \zeta_{BD} - g_{BC} \zeta_{AD} + g_{BD} \zeta_{AC} - g_{AD} \zeta_{BC},
\end{equation}
with $\left( g_{AB} \right )= \mbox{diag} \left ( k, 1,1,1 \right )$; $\zeta_{01}=\chi_1$, $\zeta_{02}=\chi_2$, $\zeta_{03}=\chi_3$, $\zeta_{12}= - \zeta_3$, $\zeta_{23}= \zeta_1$, and $\zeta_{31}=- \zeta_2$. The indices $A,B,C,D, \dots$ take the values $0,1,2,3$. The corresponding isometry group is $SO(4)$ for $k>0$ (de Sitter, in physicists' terminology), $SO(3,1)$ for $k<0$ (anti-de Sitter, in physicists' terminology), and $E(3)=SO(3)\ltimes\mathbb{R}^3$ (Euclidean) for $k=0$.

\subsubsection{Palatini Lagrangian}\label{PLFLRW}

\subsubsection*{ Half Off-Shell Potentials and Currents for Diffeomorphisms Generated by the Killing Vector Fields}
\begin{enumerate}
	\item[(i)] For the Killing vector field $\chi_1$, the potential \eqref{Udiff} acquires the form
	{\small\begin{eqnarray}
			U_{\chi_1}\!\! &=\!&2 \kappa \frac{\sqrt{1-k r^2}}{r} \cot \theta \sin \phi \, e^0\wedge e^1-2 \kappa\frac{\left(1-k r^2\right)}{r} \sin \phi \, e^0\wedge e^2 \nonumber\\
			&& - 2 \kappa \frac{\left(1-k r^2\right)}{r} \cos \theta \cos \phi \, e^0\wedge e^3 -2 \kappa \left(\frac{d a}{d t}\right) \sqrt{1-k r^2} \sin \phi \, e^1\wedge e^2 \nonumber\\
			&&-2 \kappa \left(\frac{d a}{d t}\right) \sqrt{1-k r^2} \cos \theta \cos \phi \, e^1\wedge e^3+2 \kappa \left(\frac{d a}{d t}\right) \sin \theta \cos \phi \, e^2\wedge e^3,
	\end{eqnarray}}
	and thus
	\begingroup\makeatletter\def\f@size{9}\check@mathfonts
	\def\maketag@@@#1{\hbox{\m@th\normalsize\normalfont#1}}%
	\begin{eqnarray}
		J_{\chi_1} & = & d U_{\chi_1} \nonumber\\
		& = & - 2 \kappa \frac{\sqrt{1-k r^2}}{r^2 a} \sin \phi \left[\csc ^2\theta + 2 k r^2 + 2 r^2 \left( \frac{d a}{d t}\right)^2+ r^2 a \frac{d^2 a}{d t^2} \right] e^0\wedge e^1\wedge e^2\nonumber\\
		& &-2 \kappa\frac{\sqrt{1-k r^2}}{r^2 a} \cos \phi \nonumber\\
		& & \times \left[ -\cot \theta \csc \theta+ 2 k r^2 \cos \theta + 2 r^2 \left( \frac{d a}{d t}\right)^2 \cos \theta + r^2 a \frac{d^2 a}{d t^2} \cos \theta \right] e^0\wedge e^1\wedge e^3\nonumber\\
		& &+ \frac{2 \kappa}{r^2 a} \sin \theta \cos \phi \left[-2+2 k r^2+2 r^2 \left(\frac{d a}{d t}\right)^2+r^2 a \frac{d^2 a}{ d t^2}\right] e^0\wedge e^2\wedge e^3.
	\end{eqnarray}
	\endgroup
	\item[(ii)] For the Killing vector field $\chi_2$, the potential \eqref{Udiff} becomes
	{\small\begin{eqnarray}
			U_{\chi_2} & = & -2 \kappa\frac{\sqrt{1-k r^2}}{r} \cot \theta \cos \phi \, e^0\wedge e^1+ 2 \kappa \frac{\left(1-k r^2\right)}{r} \cos \phi \, e^0\wedge e^2 \nonumber \\
			& &- 2 \kappa \frac{\left(1-k r^2\right)}{r} \cos \theta \sin \phi \, e^0\wedge e^3 +2 \kappa \left(\frac{d a}{d t}\right) \sqrt{1-k r^2} \cos \phi \, e^1\wedge e^2\nonumber\\
			& & -2 \kappa \left(\frac{d a}{d t}\right) \sqrt{1-k r^2} \cos \theta \sin \phi \, e^1\wedge e^3\!+\!2 \kappa \left(\frac{d a}{d t}\right) \sin \theta \sin \phi \, e^2\wedge e^3 , 
	\end{eqnarray}}
	and so
	\begingroup\makeatletter\def\f@size{9}\check@mathfonts
	\def\maketag@@@#1{\hbox{\m@th\normalsize\normalfont#1}}%
	\begin{eqnarray}
		J_{\chi_2} & = & d U_{\chi_2} \nonumber\\
		& = & 2 \kappa\frac{\sqrt{1-k r^2}}{r^2 a} \cos \phi \left[ \csc ^2\theta+2 k r^2 + 2 r^2 \left( \frac{d a}{d t}\right)^2 + r^2 a \frac{d^2 a}{d t^2}\right] e^0\wedge e^1\wedge e^2\nonumber\\
		& & -2 \kappa\frac{\sqrt{1-k r^2}}{r^2 a} \sin \phi \nonumber\\
		& & \times\left[ -\cot \theta \csc \theta+2 k r^2 \cos \theta +2 r^2 \left( \frac{d a}{d t}\right)^2 \cos \theta + r^2 a \frac{d^2 a}{d t^2} \cos \theta \right] e^0\wedge e^1\wedge e^3\nonumber\\
		& & +\frac{2 \kappa}{r^2 a} \sin \theta \sin \phi \left[-2+2 k r^2+2 r^2 \left(\frac{d a}{d t}\right)^2+r^2 a \frac{d^2 a}{ d t^2}\right] e^0\wedge e^2\wedge e^3.
	\end{eqnarray}
	\endgroup
	\item[(iii)] For the Killing vector field $\chi_3$, the potential \eqref{Udiff} acquires the form
	\begin{eqnarray}
		U_{\chi_3} &=&2 \kappa\frac{\left(1-k r^2\right)}{r} \sin \theta \, e^0\wedge e^3 + 2 \kappa \left(\frac{d a}{d t}\right) \sqrt{1-k r^2} \sin \theta \, e^1\wedge e^3\nonumber\\
		&&+2 \kappa \left(\frac{d a}{d t}\right) \cos \theta \, e^2\wedge e^3,
	\end{eqnarray}
	and then
	\begin{eqnarray}
		J_{\chi_3} & = & d U_{\chi_3} \nonumber\\
		& = & 2 \kappa\frac{ \sqrt{1-k r^2}}{a} \sin \theta \left[2 k+2 \left(\frac{d a}{d t}\right)^2+a \frac{d^2 a}{ d t^2}\right] e^0\wedge e^1\wedge e^3\nonumber\\
		& & +\frac{2 \kappa}{r^2 a} \cos \theta \left[-2+2 k r^2+2 r^2 \left(\frac{d a}{d t}\right)^2+r^2 a \frac{d^2 a}{ d t^2}\right] e^0\wedge e^2\wedge e^3 .
	\end{eqnarray}
	\item[(iv)] For the Killing vector field $\zeta_1$, the potential \eqref{Udiff} becomes
	\begin{eqnarray}
		U_{\zeta_1} &=&-2 \kappa \cos \theta \cot \theta \cos \phi \, e^0\wedge e^1 + 2 \kappa \sqrt{1-k r^2} \cos \theta \cos \phi \, e^0\wedge e^2 \nonumber\\
		& &-2 \kappa \sqrt{1-k r^2} \sin \phi \, e^0\wedge e^3 + 2 \kappa \left(\frac{d a}{d t}\right) r \cos \theta \cos \phi \, e^1\wedge e^2 \nonumber\\
		& &- 2 \kappa \left(\frac{d a}{d t}\right) r \sin \phi \, e^1\wedge e^3 ,
	\end{eqnarray}
	and therefore
	\begingroup\makeatletter\def\f@size{9}\check@mathfonts
	\def\maketag@@@#1{\hbox{\m@th\normalsize\normalfont#1}}%
	\begin{eqnarray}
		J_{\zeta_1} &=& d U_{\zeta_1} \nonumber\\
		& = &\frac{2 \kappa}{r a} \cos \phi \nonumber \\
		& & \times \left[\cot \theta \csc \theta + 2 k r^2 \cos \theta+2 r^2 \left(\frac{d a}{d t}\right)^2 \cos \theta+r^2 a \left(\frac{d^2 a}{ d t^2}\right) \cos \theta\right] e^0\wedge e^1\wedge e^2 \nonumber\\
		& &-\frac{2 \kappa}{r a} \sin \phi \left[-1-\cot ^2\theta+2 k r^2+2 r^2 \left(\frac{d a}{d t}\right)^2+r^2 a \frac{d^2 a}{ d t^2}\right] e^0\wedge e^1\wedge e^3.
	\end{eqnarray}
	\endgroup
	\item[(v)] For the Killing vector field $\zeta_2$, the potential \eqref{Udiff} acquires the form
	\begin{eqnarray}
		U_{\zeta_2} &=&2 \kappa \cos \theta \cot \theta \sin \phi \, e^0\wedge e^1-2 \kappa \sqrt{1-k r^2} \cos \theta \sin \phi \, e^0\wedge e^2\nonumber\\
		& & -2 \kappa \sqrt{1-k r^2} \cos \phi \, e^0\wedge e^3 - 2 \kappa \left(\frac{d a}{d t}\right) r \cos \theta \sin \phi \, e^1\wedge e^2\nonumber\\
		& &-2 \kappa \left(\frac{d a}{d t}\right) r \cos \phi \, e^1\wedge e^3,
	\end{eqnarray}
	and therefore
	\begingroup\makeatletter\def\f@size{9}\check@mathfonts
	\def\maketag@@@#1{\hbox{\m@th\normalsize\normalfont#1}}%
	\begin{eqnarray}
		J_{\zeta_2} &=& d U_{\zeta_2} \nonumber\\
		&=& -\frac{2 \kappa}{r a} \sin \phi \nonumber\\
		& &\times \left[\cot \theta \csc \theta+2 k r^2 \cos \theta+2 r^2 \left(\frac{d a}{d t}\right)^2 \cos \theta+r^2 a \left(\frac{d^2 a}{ d t^2}\right) \cos \theta\right]e^0\wedge e^1\wedge e^2 \nonumber\\
		& &-\frac{2 \kappa}{r a} \cos \phi \left[-1-\cot ^2\theta+2 k r^2+2 r^2 \left(\frac{d a}{d t}\right)^2+r^2 a \frac{d^2 a}{ d t^2}\right] e^0\wedge e^1\wedge e^3 .
	\end{eqnarray}
	\endgroup
	\item[(vi)] For the Killing vector field $\zeta_3$, the potential \eqref{Udiff} becomes
	\begin{equation}
		U_{\zeta_3} =-2 \kappa \cos \theta \, e^0\wedge e^1 + 2 \kappa \sqrt{1-k r^2} \sin \theta \, e^0\wedge e^2 + 2 \kappa \left(\frac{d a}{d t}\right) r \sin \theta \, e^1\wedge e^2,
	\end{equation}
	and therefore
	\begin{eqnarray}
		J_{\zeta_3} &=& d U_{\zeta_3} \nonumber\\
		&=&2 \kappa \frac{ r}{a} \sin \theta \left[2 k+2 \left(\frac{d a}{d t}\right)^2+a \frac{d^2 a}{ d t^2}\right]e^0\wedge e^1\wedge e^2.\label{J_3FRW}
	\end{eqnarray}
\end{enumerate}

\subsubsection*{ Half Off-Shell Potentials and Currents for Local $SO(3,1)$ Transformations Induced by the Killing Vector Fields}

\begin{enumerate}
	\item[(i)] For the gauge parameter $\tau^{IJ} \left (\chi_1 \right)$, the potential \eqref{ULor} acquires the form
	{\small\begin{equation}
			U_{\tau(\chi_1)} = 2 \kappa \frac{ \sqrt{1-k r^2}}{r} \cot \theta \sin \phi \, e^0\wedge e^1-\frac{2 \kappa}{r} \sin \phi \, e^0\wedge e^2 -\frac{2 \kappa}{r} \cos \theta \cos \phi \, e^0\wedge e^3,
	\end{equation}}
	and therefore
	\begin{eqnarray}
		J_{\tau(\chi_1)} &=& d U_{\tau(\chi_1)} \nonumber\\
		&=& -2 \kappa \frac{ \sqrt{1-k r^2} }{r^2 a} \csc ^2\theta \sin \phi \, e^0\wedge e^1\wedge e^2 \nonumber\\
		& & + 2 \kappa \frac{ \sqrt{1-k r^2}}{r^2 a} \cot \theta \csc \theta \cos \phi \, e^0\wedge e^1\wedge e^3 \nonumber\\
		& &-\frac{4 \kappa}{r^2 a} \sin \theta \cos \phi \, e^0\wedge e^2\wedge e^3.
	\end{eqnarray}
	\item[(ii)] For the gauge parameter $\tau^{IJ} \left (\chi_2 \right)$, the potential \eqref{ULor} becomes
	\begin{eqnarray}
		U_{\tau(\chi_2)} &=& -2 \kappa \frac{ \sqrt{1-k r^2}}{r} \cot \theta \cos \phi \, e^0\wedge e^1 + \frac{2 \kappa }{r} \cos \phi \, e^0\wedge e^2 \nonumber\\
		&& - \frac{2 \kappa}{r} \cos \theta \sin \phi \, e^0\wedge e^3,
	\end{eqnarray}
	and therefore
	\begin{eqnarray}
		J_{\tau(\chi_2)} &=& d U_{\tau(\chi_2)} \nonumber\\
		&=& 2 \kappa \frac{ \sqrt{1-k r^2}}{r^2 a} \csc ^2\theta \cos \phi \, e^0\wedge e^1\wedge e^2\nonumber\\
		& &+ 2 \kappa \frac{\sqrt{1-k r^2} }{r^2 a} \cot \theta \csc \theta \sin \phi \, e^0\wedge e^1\wedge e^3\nonumber\\
		& &-\frac{4 \kappa}{r^2 a} \sin \theta \sin \phi \, e^0\wedge e^2\wedge e^3.
	\end{eqnarray}
	\item[(iii)] For the gauge parameter $\tau^{IJ} \left (\chi_3 \right)$, the potential \eqref{ULor} acquires the form
	\begin{equation}
		U_{\tau(\chi_3)} =\frac{2 \kappa}{r} \sin \theta \, e^0\wedge e^3,
	\end{equation}
	and therefore
	\begin{eqnarray}
		J_{\tau(\chi_3)} &=& d U_{\tau(\chi_3)} \nonumber\\
		&=&-\frac{4 \kappa}{r^2 a} \cos \theta \, e^0\wedge e^2\wedge e^3.
	\end{eqnarray}
	\item[(iv)] For the gauge parameter $\tau^{IJ} \left (\zeta_1 \right)$, the potential \eqref{ULor} becomes
	\begin{equation}
		U_{\tau(\zeta_1)} = -2 \kappa \csc \theta \cos \phi \, e^0 \wedge e^1,
	\end{equation}
	and therefore
	\begin{eqnarray}
		J_{\tau(\zeta_1)} &=& d U_{\tau(\zeta_1)} \nonumber\\
		&=& \frac{2 \kappa }{r a} \cot \theta \csc \theta \cos \phi \, e^0\wedge e^1\wedge e^2 + \frac{2 \kappa }{r a} \csc ^2\theta \sin \phi \, e^0\wedge e^1\wedge e^3. 
	\end{eqnarray}
	\item[(v)] For the gauge parameter $\tau^{IJ} \left (\zeta_2 \right)$, the potential \eqref{ULor} acquires the form
	
	\begin{equation}
		U_{\tau(\zeta_2)} =2 \kappa \csc \theta \sin \phi \, e^0\wedge e^1,
	\end{equation}
	and therefore
	\begin{eqnarray}
		J_{\tau(\zeta_2)} &=& d U_{\tau(\zeta_2)} \nonumber\\
		&=&-\frac{2 \kappa}{r a} \cot \theta \csc \theta \sin \phi \, e^0\wedge e^1\wedge e^2+\frac{2 \kappa}{r a} \csc ^2\theta \cos \phi \, e^0\wedge e^1\wedge e^3.
	\end{eqnarray}
	\item[(vi)] For the gauge parameter $\tau^{IJ} \left (\zeta_3 \right)$, the potential \eqref{ULor} and its current become
	\begin{eqnarray}
		U_{\tau(\zeta_3)} = 0, \quad J_{\tau(\zeta_3)} = 0.
	\end{eqnarray}
	This is so because $\tau^{IJ} \left ( \zeta_3 \right )$ vanishes.
\end{enumerate}

\subsubsection{Holst Lagrangian}

\subsubsection*{ Half Off-Shell Potentials and Currents for Diffeomorphisms Generated by the Killing Vector Fields}
\begin{enumerate}
	\item[(i)] For the Killing vector field $\chi_1$, the potential \eqref{UHvsUP2} acquires the form
	{\small\begin{eqnarray}
			U_{\chi_1} &\!\!=\!\!& 2 \kappa\frac{\sqrt{1-k r^2}}{r} \cot \theta \sin \phi \, e^0\wedge e^1
			-2 \kappa\frac{\left(1-k r^2\right)}{r} \sin \phi \, e^0\wedge e^2 \nonumber\\
			& &-2 \kappa\frac{\left(1-k r^2\right)}{r} \cos \theta \cos \phi \, e^0\wedge e^3 - 2 \kappa \left( \frac{d a}{d t}\right) \sqrt{1-k r^2} \sin \phi \, e^1\wedge e^2
			\nonumber\\
			& & -2 \kappa \left( \frac{d a}{d t}\right) \sqrt{1-k r^2} \cos \theta \cos \phi \, e^1\wedge e^3 + 2 \kappa \left( \frac{d a}{d t}\right) \sin \theta \cos \phi \, e^2\wedge e^3 \nonumber\\
			&&- \frac{2 \kappa }{\gamma r} \sqrt{1-k r^2} \cot \theta \sin \phi \, e^2\wedge e^3 
			-\frac{2 \kappa}{\gamma r} \left(1-k r^2\right)\sin \phi \, e^1\wedge e^3\nonumber\\
			& & +\frac{2 \kappa}{\gamma r} \left(1-k r^2\right) \cos \theta \cos \phi \, e^1\wedge e^2
			-\frac{2 \kappa}{\gamma } \left( \frac{d a}{d t}\right) \sqrt{1-k r^2} \sin \phi \, e^0\wedge e^3\nonumber\\
			& & + \frac{2 \kappa}{\gamma } \! \left( \frac{d a}{d t}\right) \! \sqrt{1-k r^2} \cos \theta \cos \phi \, e^0\wedge e^2
			+\frac{2 \kappa}{\gamma } \left( \frac{d a}{d t}\right) \sin \theta \cos \phi \, e^0\wedge e^1,
	\end{eqnarray}}
	and thus
	\begin{eqnarray}
		J_{\chi_1} & = & d U_{\chi_1} \nonumber\\
		& = & \Biggl\{ -2 \kappa\frac{\sqrt{1-k r^2}}{r^2 a} \sin \phi \left[\csc ^2\theta + 2 k r^2 + 2 r^2 \left( \frac{d a}{d t}\right)^2+ r^2 a \frac{d^2 a}{d t^2} \right] \nonumber\\
		& & + \frac{4 \kappa}{\gamma r a}\left( \frac{d a}{d t}\right) \cos \theta \cos \phi \Biggr\} e^0\wedge e^1\wedge e^2 + \Biggl\{ -2 \kappa \frac{ \sqrt{1-k r^2}}{r^2 a} \cos \phi \nonumber\\
		& & \times \left[ -\cot \theta \csc \theta+ 2 k r^2 \cos \theta + 2 r^2 \left( \frac{d a}{d t}\right)^2 \cos \theta + r^2 a \frac{d^2 a}{d t^2} \cos \theta \right] \nonumber\\
		& & -\frac{4 \kappa }{ \gamma r a} \left( \frac{d a}{d t}\right) \sin \phi \Biggr\} e^0\wedge e^1\wedge e^3 \nonumber\\
		& &+\Biggl\{ \frac{2 \kappa}{r^2 a} \sin \theta \cos \phi \left[-2+2 k r^2 + 2 r^2 \left( \frac{d a}{d t}\right)^2 + r^2 a \frac{d^2 a}{d t^2} \right] \nonumber\\
		& & -\frac{4 \kappa}{\gamma r a}\left( \frac{d a}{d t}\right) \sqrt{1-k r^2} \cot \theta \sin \phi\Biggr\} e^0\wedge e^2\wedge e^3 \nonumber\\
		&& -\frac{2 \kappa}{\gamma r^2 a} \left(1-2 k r^2\right) \cot \theta \sin \phi \, e^1\wedge e^2\wedge e^3.
	\end{eqnarray}
	\item[(ii)] For the Killing vector field $\chi_2$, the potential \eqref{UHvsUP2} becomes
	{\small\begin{eqnarray}
			U_{\chi_2} &\!\! = \!\! & -2 \kappa \frac{\sqrt{1-k r^2}}{r} \cot \theta \cos \phi \, e^0\wedge e^1
			+2 \kappa \frac{\left(1-k r^2\right)}{r} \cos \phi e^0\wedge \, e^2\nonumber\\
			& & -2 \kappa \frac{\left(1-k r^2 \right)}{r} \cos \theta \sin \phi \, e^0\wedge e^3
			+2 \kappa \left( \frac{d a}{d t}\right) \sqrt{1-k r^2} \cos \phi \, e^1\wedge e^2\nonumber\\
			& & -2 \kappa \left( \frac{d a}{d t}\right) \sqrt{1-k r^2} \cos \theta \sin \phi \, e^1\wedge e^3
			+2 \kappa \left( \frac{d a}{d t}\right) \sin \theta \sin \phi \, e^2\wedge e^3 \nonumber\\
			& & +\frac{2 \kappa}{\gamma r} \sqrt{1-k r^2} \cot \theta \cos \phi \, e^2\wedge e^3
			+\frac{2 \kappa}{\gamma r} \left(1-k r^2\right) \cos \phi e^1\wedge \, e^3\nonumber\\
			& & +\frac{2 \kappa}{\gamma r} \left(1-k r^2\right) \cos \theta \sin \phi \, e^1\wedge e^2
			+\frac{2 \kappa}{\gamma }\left( \frac{d a}{d t}\right) \sqrt{1-k r^2} \cos \phi \, e^0\wedge e^3\nonumber\\
			& & +\frac{2 \kappa}{\gamma }\!\left( \frac{d a}{d t}\right) \sqrt{1-k r^2} \cos \theta \sin \phi \, e^0\wedge e^2
			+\frac{2 \kappa}{\gamma } \left( \frac{d a}{d t}\right) \sin \theta \sin \phi \, e^0\wedge e^1,
	\end{eqnarray}}
	and so
	\begin{eqnarray}
		J_{\chi_2} & = & d U_{\chi_2} \nonumber\\
		& = & \Biggl\{ 2 \kappa \frac{\sqrt{1-k r^2}}{r^2 a} \cos \phi \left[ \csc ^2\theta+2 k r^2 + 2 r^2 \left( \frac{d a}{d t}\right)^2 + r^2 a \frac{d^2 a}{d t^2}\right] \nonumber\\
		& & + \frac{4 \kappa}{\gamma r a } \left( \frac{d a}{d t}\right) \cos \theta \sin \phi \Biggr\} e^0\wedge e^1\wedge e^2 + \Biggl\{-2 \kappa \frac{\sqrt{1-k r^2}}{r^2 a} \sin \phi\nonumber\\
		& &\times \left[ -\cot \theta \csc \theta+2 k r^2 \cos \theta +2 r^2 \left( \frac{d a}{d t}\right)^2 \cos \theta + r^2 a \frac{d^2 a}{d t^2} \cos \theta \right] \nonumber\\
		& & + \frac{4 \kappa}{\gamma r a}\left( \frac{d a}{d t}\right) \cos \phi \Biggr\} e^0\wedge e^1\wedge e^3 \nonumber \\
		& & + \Biggl\{ \frac{2 \kappa }{r^2 a} \sin \theta \sin \phi \left[-2 + 2 k r^2 + 2 r^2 \left( \frac{d a}{d t}\right)^2 + r^2 a \frac{d^2 a}{d t^2}\right] \nonumber\\
		& & + \frac{4 \kappa}{ \gamma r a} \left( \frac{d a}{d t}\right) \sqrt{1-k r^2} \cot \theta \cos \phi \Biggr\} e^0\wedge e^2\wedge e^3 \nonumber\\
		& & +\frac{2 \kappa}{ \gamma r^2 a} \left(1-2 k r^2\right) \cot \theta \cos \phi \, e^1\wedge e^2\wedge e^3. 
	\end{eqnarray}
	\item[(iii)] For the Killing vector field $\chi_3$, the potential \eqref{UHvsUP2} acquires the form
	\begin{eqnarray}
		U_{\chi_3} &=& 2 \kappa \frac{\left(1-k r^2\right)}{r} \sin \theta \, e^0\wedge e^3 + 2 \kappa \left( \frac{d a}{d t}\right) \sqrt{1-k r^2} \sin \theta \, e^1\wedge e^3 \nonumber\\
		& & + 2 \kappa \left( \frac{d a}{d t}\right) \cos \theta \, e^2\wedge e^3 - \frac{2 \kappa}{\gamma r} \left(1-k r^2\right) \sin \theta \, e^1\wedge e^2 \nonumber\\
		&& - \frac{2 \kappa}{\gamma } \left( \frac{d a}{d t}\right) \sqrt{1-k r^2} \sin \theta \, e^0\wedge e^2 + \frac{2 \kappa}{\gamma } \left( \frac{d a}{d t}\right) \cos \theta \, e^0\wedge e^1, 
	\end{eqnarray}
	and then
	\begin{eqnarray}
		J_{\chi_3} & = & d U_{\chi_3} \nonumber\\
		&=& -\frac{4 \kappa}{a \gamma r} \left( \frac{d a}{d t}\right) \sin \theta \, e^0\wedge e^1\wedge e^2 \nonumber \\
		&&+2 \kappa \frac{\sqrt{1-k r^2}}{a} \sin \theta \left[2 k + 2 \left( \frac{d a}{d t}\right)^2+a \frac{d^2 a}{d t^2}\right] e^0\wedge e^1\wedge e^3 \nonumber\\
		& & +\frac{2 \kappa}{r^2 a} \cos \theta \left[ - 2 + 2 k r^2 + 2 r^2 \left( \frac{d a}{d t}\right)^2 + r^2 a \frac{d^2 a}{d t^2} \right]e^0\wedge e^2\wedge e^3. 
	\end{eqnarray}
	\item[(iv)] For the Killing vector field $\zeta_1$, the potential \eqref{UHvsUP2} becomes
	\begin{eqnarray}
		U_{\zeta_1} &=& -2 \kappa \cos \theta \cot \theta \cos \phi \, e^0\wedge e^1 + 2 \kappa \sqrt{1-k r^2} \cos \theta \cos \phi \, e^0\wedge e^2 \nonumber\\
		& & -2 \kappa \sqrt{1-k r^2} \sin \phi \, e^0\wedge e^3 + 2 \kappa \left( \frac{d a}{d t}\right) r \cos \theta \cos \phi \, e^1\wedge e^2 \nonumber\\
		& &- 2 \kappa \left( \frac{d a}{d t}\right) r \sin \phi \, e^1\wedge e^3 +\frac{2 \kappa}{\gamma } \cos \theta \cot \theta \cos \phi \, e^2\wedge e^3 \nonumber\\
		& &+\frac{2 \kappa}{\gamma } \sqrt{1-k r^2} \cos \theta \cos \phi \, e^1\wedge e^3 +\frac{2 \kappa}{\gamma } \sqrt{1-k r^2} \sin \phi \, e^1\wedge e^2
		\nonumber\\
		& & +\frac{2 \kappa}{\gamma } \left( \frac{d a}{d t}\right) r \cos \theta \cos \phi \, e^0\wedge e^3 + \frac{2 \kappa}{\gamma } \left( \frac{d a}{d t}\right) r \sin \phi \, e^0\wedge e^2, 
	\end{eqnarray}
	and therefore
	{\small\begin{eqnarray}
			J_{\zeta_1} &=& d U_{\zeta_1} \nonumber\\
			&=& \frac{2 \kappa}{r a} \cos \phi \left[ \cot \theta \csc \theta + 2 k r^2 \cos \theta + 2 r^2 \left( \frac{d a}{d t}\right)^2 \cos \theta + r^2 a \frac{d^2 a}{d t^2} \cos \theta \right] e^0\wedge e^1\wedge e^2 \nonumber\\
			& & -\frac{2 \kappa}{r a} \sin \phi \left[- 1 -\cot ^2\theta + 2 k r^2 + 2 r^2 \left( \frac{d a}{d t}\right)^2 + r^2 a \frac{d^2 a}{d t^2} \right] e^0\wedge e^1\wedge e^3\nonumber\\
			& & +\frac{4 \kappa}{\gamma a} \left( \frac{d a}{d t}\right) \csc \theta \cos \phi \, e^0\wedge e^2\wedge e^3 +\frac{4 \kappa}{\gamma r a } \sqrt{1-k r^2} \csc \theta \cos \phi \, e^1\wedge e^2\wedge e^3.
	\end{eqnarray}}
	\item[(v)] For the Killing vector field $\zeta_2$, the potential \eqref{UHvsUP2} acquires the form
	\begin{eqnarray}
		U_{\zeta_2} &=& 2 \kappa \cos \theta \cot \theta \sin \phi \, e^0\wedge e^1
		-2 \kappa \sqrt{1-k r^2} \cos \theta \sin \phi \, e^0\wedge e^2\nonumber\\
		& & -2 \kappa \sqrt{1-k r^2} \cos \phi \, e^0\wedge e^3 -2 \kappa \left( \frac{d a}{d t}\right) r \cos \theta \sin \phi \, e^1\wedge e^2 \nonumber\\
		& & -2 \kappa \left( \frac{d a}{d t}\right) r \cos \phi \, e^1\wedge e^3 -\frac{2 \kappa}{\gamma } \cos \theta \cot \theta \sin \phi \, e^2\wedge e^3
		\nonumber\\
		& & -\frac{2 \kappa}{\gamma } \sqrt{1-k r^2} \cos \theta \sin \phi \, e^1\wedge e^3 +\frac{2 \kappa}{\gamma } \sqrt{1-k r^2} \cos \phi \, e^1\wedge e^2 \nonumber\\
		& & 	-\frac{2 \kappa}{\gamma } \left( \frac{d a}{d t}\right) r \cos \theta \sin \phi \, e^0\wedge e^3 + \frac{2 \kappa}{\gamma } \left( \frac{d a}{d t}\right) r \cos \phi \, e^0\wedge e^2, 
	\end{eqnarray}
	and therefore
	{\small\begin{eqnarray}
			J_{\zeta_2} &=& d U_{\zeta_2} \nonumber\\
			&=& - \frac{2 \kappa}{r a} \sin \phi \nonumber\\
			& & \times \left[ \cot \theta \csc \theta + 2 k r^2 \cos \theta + 2 r^2 \left( \frac{d a}{d t}\right)^2 \cos \theta + r^2 a \frac{d^2 a}{d t^2} \cos \theta \right] e^0\wedge e^1\wedge e^2\nonumber\\
			& & -\frac{2 \kappa}{r a} \cos \phi \left[ - 1 -\cot^2 \theta + 2 k r^2 + 2 r^2 \left( \frac{d a}{d t}\right)^2 + r^2 a \frac{d^2 a}{d t^2} \right] e^0\wedge e^1\wedge e^3\nonumber\\
			& & -\frac{4 \kappa }{\gamma a} \left( \frac{d a}{d t}\right) \csc \theta \sin \phi \, e^0\wedge e^2 \wedge e^3 -\frac{4 \kappa}{\gamma r a } \sqrt{1-k r^2} \csc \theta \sin \phi \, e^1\wedge e^2\wedge e^3.
	\end{eqnarray}}
	\item[(vi)] For the Killing vector field $\zeta_3$, the potential \eqref{UHvsUP2} becomes
	{\small\begin{eqnarray}\label{UFRWzeta3}
			U_{\zeta_3} &=& -2 \kappa \cos \theta \, e^0\wedge e^1 + 2 \kappa \sqrt{1-k r^2} \sin \theta \, e^0\wedge e^2 + 2 \kappa \left( \frac{d a}{d t}\right) r \sin \theta \, e^1\wedge e^2\nonumber\\
			& & + \frac{2 \kappa}{\gamma } \cos \theta \, e^2 \wedge e^3 +\frac{2 \kappa}{\gamma } \sqrt{1-k r^2} \sin \theta \, e^1\wedge e^3 + \frac{2 \kappa }{\gamma } \left( \frac{d a}{d t}\right) r \sin \theta \, e^0\!\wedge e^3,
	\end{eqnarray}}
	and therefore
	\begin{eqnarray}
		J_{\zeta_3} &=& d U_{\zeta_3} \nonumber\\
		&=& 2 \kappa \frac{r}{a} \sin \theta \left[ 2 k + 2 \left( \frac{d a}{d t}\right)^2 +a \frac{d^2 a}{d t^2} \right] e^0\wedge e^1\wedge e^2.\label{JFRWzeta3}
	\end{eqnarray}
\end{enumerate}

Notice that the potentials and currents that depend on the Immirzi parameter will still depend on it even in the `on-shell' case, except for the current \eqref{JFRWzeta3}, which actually coincides with the current \eqref{J_3FRW} for the Palatini Lagrangian. This is a consequence of the fact that the terms on the second row of the potential \eqref{UFRWzeta3} (that involve the Immirzi parameter) can be cast as the exact form $d(\kappa\gamma^{-1} a r \sin\theta\ e^3)$ and hence do not contribute to the current \eqref{JFRWzeta3}.

\subsubsection*{ Half Off-Shell Potentials and Currents for Local $SO(3,1)$ Transformations Induced by the Killing Vector Fields}
\begin{enumerate}
	\item[(i)] For the gauge parameter $\tau^{IJ} \left ( \chi_1 \right )$, the potential \eqref{ULorH} acquires the form
	{\footnotesize\begin{eqnarray}
			U_{\tau(\chi_1)} &=& 2 \kappa \frac{ \sqrt{1-k r^2} }{r} \cot \theta \sin \phi \, e^0\wedge e^1-\frac{2 \kappa}{r} \sin \phi \, e^0\wedge e^2-\frac{2 \kappa}{r} \cos \theta \cos \phi \, e^0\wedge e^3 \nonumber \\
			& &-\frac{2 \kappa }{\gamma r} \sqrt{1-k r^2} \cot \theta \sin \phi \, e^2\wedge e^3-\frac{2 \kappa}{\gamma r} \sin \phi \, e^1\wedge e^3 + \frac{2 \kappa}{\gamma r} \cos \theta \cos \phi \, e^1\wedge e^2, 
	\end{eqnarray}}
	and therefore
	\begin{eqnarray}
		J_{\tau(\chi_1)} &=& d U_{\tau(\chi_1)} \nonumber\\
		&=& \left[ -2 \kappa \frac{\sqrt{1-k r^2}}{r^2 a} \csc ^2\theta \sin \phi + \frac{4 \kappa}{\gamma r a} \left( \frac{d a}{d t}\right) \cos \theta \cos \phi \right] e^0\wedge e^1\wedge e^2\nonumber\\
		& & + \left[ 2 \kappa \frac{\sqrt{1-k r^2}}{r^2 a} \cot \theta \csc \theta \cos \phi-\frac{4 \kappa}{ \gamma r a} \left( \frac{d a}{d t}\right) \sin \phi \right] e^0\wedge e^1\wedge e^3\nonumber\\
		& & + \left[ - \frac{4 \kappa}{r^2 a} \sin \theta \cos \phi - \frac{4 \kappa}{\gamma r a}\left( \frac{d a}{d t}\right) \sqrt{1-k r^2} \cot \theta \sin \phi \right] e^0\wedge e^2\wedge e^3\nonumber\\
		& & -\frac{2 \kappa} {\gamma r^2 a} \left(1-2 k r^2\right) \cot \theta \sin \phi \, e^1\wedge e^2\wedge e^3.
	\end{eqnarray}
	\item[(ii)] For the gauge parameter $\tau^{IJ} \left ( \chi_2 \right )$, the potential \eqref{ULorH} becomes
	{\footnotesize\begin{eqnarray}
			U_{\tau(\chi_2)} &=&-2 \kappa \frac{\sqrt{1-k r^2}}{r} \cot \theta \cos \phi \, e^0\wedge e^1+\frac{2 \kappa}{r} \cos \phi \, e^0\wedge e^2 -\frac{2 \kappa}{r} \cos \theta \sin \phi \, e^0\wedge e^3\nonumber\\
			& &+\frac{2 \kappa}{\gamma r} \sqrt{1-k r^2} \cot \theta \cos \phi \, e^2\wedge e^3 + \frac{2 \kappa}{\gamma r} \cos \phi \, e^1\wedge e^3+\frac{2 \kappa}{\gamma r} \cos \theta \sin \phi \, e^1\wedge e^2, 
	\end{eqnarray}}
	and therefore
	\begin{eqnarray}
		J_{\tau(\chi_2)} &=& d U_{\tau(\chi_2)} \nonumber\\
		&=& \left[ 2 \kappa \frac{ \sqrt{1-k r^2} }{r^2 a} \csc ^2\theta \cos \phi + \frac{4 \kappa}{\gamma r a} \left( \frac{d a}{d t}\right) \cos \theta \sin \phi \right] e^0\wedge e^1\wedge e^2\nonumber\\
		& & + \left[ 2 \kappa \frac{\sqrt{1-k r^2}}{r^2 a} \cot \theta \csc \theta \sin \phi+\frac{4 \kappa}{\gamma r a} \left( \frac{d a}{d t}\right) \cos \phi \right] e^0\wedge e^1\wedge e^3\nonumber\\
		& & + \left[-\frac{4 \kappa}{r^2 a} \sin \theta \sin \phi + \frac{4 \kappa}{\gamma r a} \left( \frac{d a}{d t}\right) \sqrt{1-k r^2} \cot \theta \cos \phi \right]e^0\wedge e^2\wedge e^3\nonumber\\
		& & +\frac{2 \kappa}{ \gamma r^2 a} \left(1-2 k r^2\right) \cot \theta \cos \phi \, e^1\wedge e^2\wedge e^3.
	\end{eqnarray}
	\item[(iii)] For the gauge parameter $\tau^{IJ} \left ( \chi_3 \right )$, the potential \eqref{ULorH} acquires the form
	\begin{equation}
		U_{\tau(\chi_3)} =\frac{2 \kappa}{r} \sin \theta \, e^0\wedge e^3-\frac{2 \kappa}{\gamma r} \sin \theta \, e^1\wedge e^2,
	\end{equation}
	and therefore
	\begin{eqnarray}
		J_{\tau(\chi_3)} &=& d U_{\tau(\chi_3)} \nonumber\\
		&=&-\frac{4 \kappa }{r^2 a}\cos \theta \, e^0\wedge e^2\wedge e^3-\frac{4 \kappa}{\gamma r a}\left( \frac{d a}{d t}\right) \sin \theta \, 
		e^0\wedge e^1\wedge e^2.
	\end{eqnarray}
	\item[(iv)] For the gauge parameter $\tau^{IJ} \left ( \zeta_1 \right )$, the potential \eqref{ULorH} becomes
	\begin{equation}
		U_{\tau(\zeta_1)} = -2 \kappa \csc \theta \cos \phi \, e^0\wedge e^1 + \frac{2 \kappa}{\gamma } \csc \theta \cos \phi \, e^2\wedge e^3,
	\end{equation}
	and therefore
	\begin{eqnarray}
		J_{\tau(\zeta_1)} &=& d U_{\tau(\zeta_1)} \nonumber\\
		&=& \frac{2 \kappa}{ r a} \cot \theta \csc \theta \cos \phi \, e^0\wedge e^1\wedge e^2
		+ \frac{2 \kappa}{ r a} \csc ^2\theta \sin \phi \, e^0\wedge e^1\wedge e^3\nonumber\\
		& & +\frac{4 \kappa}{ \gamma a} \left( \frac{d a}{d t}\right) \csc \theta \cos \phi \, e^0\wedge e^2\wedge e^3 +\frac{4 \kappa }{ \gamma r a} \sqrt{1-k r^2}\csc \theta \cos \phi \, e^1\wedge e^2\wedge e^3.
		\nonumber\\
		& &
	\end{eqnarray}
	\item[(v)] For the gauge parameter $\tau^{IJ} \left ( \zeta_2 \right )$, the potential \eqref{ULorH} acquires the form
	\begin{equation}
		U_{\tau(\zeta_2)} =2 \kappa \csc \theta \sin \phi \, e^0\wedge e^1 -\frac{2 \kappa}{\gamma } \csc \theta \sin \phi \, e^2\wedge e^3,
	\end{equation}
	and therefore
	{\small\begin{eqnarray}
			J_{\tau(\zeta_2)} &=& d U_{\tau(\zeta_2)} \nonumber\\
			&=& -\frac{2 \kappa }{r a} \cot \theta \csc \theta \sin \phi \, e^0\wedge e^1\wedge e^2+\frac{2 \kappa}{r a} \csc ^2\theta \cos \phi \, 
			e^0\wedge e^1\wedge e^3\nonumber\\
			& & -\frac{4 \kappa}{\gamma a}\left( \frac{d a}{d t}\right) \csc \theta \sin \phi \, e^0\wedge e^2\wedge e^3 -\frac{4 \kappa}{\gamma r a} \sqrt{1-k r^2} \csc \theta \sin \phi \, e^1\wedge e^2\wedge e^3. 
	\end{eqnarray}}
	\item[(vi)] For the gauge parameter $\tau^{IJ} \left ( \zeta_3 \right )$, the potential \eqref{ULorH} and its current become
	\begin{eqnarray}
		U_{\tau(\zeta_3)} =0, \quad J_{\tau(\zeta_3)}= 0.
	\end{eqnarray}
	This is so because $\tau^{IJ} \left ( \zeta_3 \right )$ vanishes.
\end{enumerate}

In this case,  all the nonvanishing potentials and currents depend on the Immirzi parameter, which also holds in the `on-shell' case.

\subsubsection*{ Half Off-Shell Potentials and Currents \eqref{UHon} and \eqref{JHon}}
\begin{enumerate}
	\item[(i)] For $\chi_1$
	{\footnotesize\begin{eqnarray}
			U_H &=& U_{\chi_1}- U_{\tau \left ( \chi_1 \right )} \nonumber\\
			&=&U_P+d\left[ \frac{\kappa }{\gamma} ( \chi_1 \intprod e_I) e^I \right] \nonumber\\
			&=&2 \kappa k r \sin\phi e^0\wedge e^2 +2 \kappa k r \cos\theta \cos\phi e^0\wedge e^3 -2\kappa \left(\frac{d a}{d t}\right) \sqrt{1-k r^2} \sin\phi \, e^1\wedge e^2 \nonumber\\
			&& -2 \kappa \left(\frac{d a}{d t}\right) \sqrt{1-k r^2} \cos\theta \cos\phi \, e^1\wedge e^3 +2 \kappa \left(\frac{d a}{d t}\right) \sin\theta \cos\phi \, e^2\wedge e^3\nonumber\\
			&&+d\left( \frac{ \kappa a}{\gamma } \sin\theta \cos\phi \, e^1 + \frac{ \kappa a}{\gamma } \sqrt{1-k r^2} \cos\theta \cos\phi \, e^2 - \frac{ \kappa a}{\gamma } \sqrt{1-k r^2} \sin\phi \, e^3 \right), \label{FLRW-UH-1} \nonumber\\
			&&\\
			J_H &=& = d U_H = J_P = J_{\chi_1}- J_{\tau \left ( \chi_1 \right )}\nonumber \\
			& = & -\frac{2 \kappa }{a} \sqrt{1-k r^2} \sin\phi \left[2 k +2 \left(\frac{d a}{d t}\right)^2 + a \frac{d^2 a}{ d t^2} \right] e^0\wedge e^1\wedge e^2 \nonumber \\
			&&-\frac{2 \kappa }{a} \sqrt{1-k r^2} \cos\theta \cos\phi \left[2 k +2 \left(\frac{d a}{d t}\right)^2 +a \frac{d^2 a}{ d t^2} \right] e^0\wedge e^1\wedge e^3\nonumber\\
			&&+\frac{2 \kappa }{a} \sin\theta \cos\phi \left[2 k +2 \left(\frac{d a}{d t}\right)^2 + a \frac{d^2 a}{ d t^2}\right] e^0\wedge e^2\wedge e^3.
	\end{eqnarray}}
	Notice that $U_P$ is given by the first five terms in the last equality in \eqref{FLRW-UH-1}.\\
	\item[(ii)] For $\chi_2$
	{\small\begin{eqnarray}
			U_H &=& U_{\chi_2}- U_{\tau \left ( \chi_2 \right )} \nonumber\\
			&=&U_P+d\left[ \frac{\kappa }{\gamma} ( \chi_2 \intprod e_I) e^I \right] \nonumber\\
			&=& -2 \kappa k r \cos\phi \, e^0\wedge e^2 +2 \kappa k r \cos\theta \sin\phi\, e^0\wedge e^3 +2\kappa \left(\frac{d a}{d t}\right) \sqrt{1-k r^2} \cos\phi \, e^1\wedge e^2 \nonumber\\
			&&-2\kappa \left(\frac{d a}{d t}\right) \sqrt{1-k r^2} \cos\theta \sin\phi \, e^1\wedge e^3+2\kappa \left(\frac{d a}{d t}\right) \sin\theta \sin\phi \, e^2\wedge e^3 \nonumber\\
			&&+d\left( \frac{ \kappa a}{\gamma } \sin\theta \sin\phi \, e^1+\frac{ \kappa a}{\gamma } \sqrt{1-k r^2} \cos\theta \sin\phi \, e^2+\frac{ \kappa a}{\gamma } \sqrt{1-k r^2} \cos\phi \, e^3 \right), \nonumber\\
			&&\label{FLRW-UH-2}\\
			J_H &=& = d U_H = J_P = J_{\chi_2}- J_{\tau \left ( \chi_2 \right )}\nonumber \\
			& = &\frac{2 \kappa }{a} \sqrt{1-k r^2} \cos\phi \left[2 k+2 \left(\frac{d a}{d t}\right)^2+a \frac{d^2 a}{ d t^2}\right] e^0\wedge e^1\wedge e^2 \nonumber\\
			&& -\frac{2 \kappa}{a} \sqrt{1-k r^2} \cos\theta \sin\phi \left[2 k+2 \left(\frac{d a}{d t}\right)^2+a \frac{d^2 a}{ d t^2}\right] e^0\wedge e^1\wedge e^3\nonumber\\
			&&+\frac{2 \kappa }{a} \sin\theta \sin\phi \left[2 k+2 \left(\frac{d a}{d t}\right)^2+a \frac{d^2 a}{ d t^2}\right] e^0\wedge e^2\wedge e^3 .
	\end{eqnarray}}
	Notice that $U_P$ is given by the first five terms in the last equality in \eqref{FLRW-UH-2}.\\
	\item[(iii)] For $\chi_3$
	\begin{eqnarray}
		U_H &=& U_{\chi_3}- U_{\tau \left ( \chi_3 \right )} \nonumber\\
		&=&U_P+d\left[ \frac{\kappa }{\gamma} ( \chi_3 \intprod e_I) e^I \right] \nonumber\\
		&=& -2 \kappa k r \sin\theta\, e^0\wedge e^3 + 2\kappa \left(\frac{d a}{d t}\right) \sqrt{1-k r^2} \sin\theta \, e^1\wedge e^3+2\kappa \left(\frac{d a}{d t}\right) \cos\theta \, e^2\wedge e^3 \nonumber\\
		&&+d\left( \frac{ \kappa a}{\gamma } \cos\theta \, e^1-\frac{\kappa a }{\gamma } \sqrt{1-k r^2} \sin\theta \, e^2 \right), \label{FLRW-UH-3}\\
		J_H &=& d U_H = J_P = J_{\chi_3}- J_{\tau \left ( \chi_3 \right )}\nonumber \\
		& = & \frac{2 \kappa }{a} \sqrt{1-k r^2} \sin\theta \left[ 2 k+2\left(\frac{d a}{d t}\right)^2+a \frac{d^2 a}{ d t^2}\right] e^0\wedge e^1\wedge e^3 \nonumber\\
		& & +\frac{2 \kappa }{a} \cos\theta \left[2 k+2 \left(\frac{d a}{d t}\right)^2 + a \frac{d^2 a}{ d t^2}\right] e^0\wedge e^2\wedge e^3.
	\end{eqnarray}
	Notice that $U_P$ is given by the first three terms of the last equality in \eqref{FLRW-UH-3}.\\
	\item[(iv)] For $\zeta_1$
	{\small\begin{eqnarray}
			U_H &=& U_{\zeta_1}- U_{\tau \left ( \zeta_1 \right )} \nonumber\\
			&=&U_P+d\left[ \frac{\kappa }{\gamma} ( \zeta_1 \intprod e_I) e^I \right] \nonumber\\
			&=& 2 \kappa \sin\theta \cos\phi e^0\wedge e^1 +2 \kappa \sqrt{1-k r^2} \cos\theta \cos\phi \, e^0\wedge e^2 -2 \kappa \sqrt{1-k r^2} \sin\phi e^0\wedge e^3 \nonumber\\
			&& + 2\kappa \left(\frac{d a}{d t}\right) r \cos\theta \cos\phi \, e^1\wedge e^2-2 \kappa \left(\frac{d a}{d t}\right) r \sin\phi \, e^1\wedge e^3 \nonumber\\
			&&+d\left(\frac{ \kappa a r }{\gamma } \sin\phi \, e^2 + \frac{ \kappa a r}{\gamma } \cos\theta \cos\phi \, e^3 \right), \label{FLRW-UH-4}\\
			J_H &=& d U_H = J_P = J_{\zeta_1}- J_{\tau \left ( \zeta_1 \right )}\nonumber \\
			& = & 2 \kappa \frac{ r }{a} \cos\theta \cos\phi \left[2 k+2 \left(\frac{d a}{d t}\right)^2+a \frac{d^2 a}{ d t^2}\right] e^0\wedge e^1\wedge e^2 \nonumber\\
			& & -2 \kappa \frac{ r }{a} \sin\phi \left[2 k+2 \left(\frac{d a}{d t}\right)^2+a \frac{d^2 a}{ d t^2}\right] e^0\wedge e^1\wedge e^3. 
	\end{eqnarray}}
	Notice that $U_P$ is given by the first five terms in the last equality in \eqref{FLRW-UH-4}.
	\item[(v)] For $\zeta_2$
	\begin{eqnarray}
		U_H &=& U_{\zeta_2}- U_{\tau \left ( \zeta_2 \right )} \nonumber\\
		&=&U_P+d\left[ \frac{\kappa }{\gamma} ( \zeta_2 \intprod e_I) e^I \right] \nonumber\\
		&=& -2 \kappa \sin\theta \sin\phi \, e^0\wedge e^1 -2 \kappa \sqrt{1-k r^2} \cos\phi \, e^0\wedge e^3 \nonumber\\
		& & -2 \kappa \sqrt{1-k r^2} \cos\theta \sin\phi \, e^0\wedge e^2 -2\kappa \left(\frac{d a}{d t}\right) r \cos\theta \sin\phi \, e^1\wedge e^2\nonumber\\
		&&-2 \kappa \left(\frac{d a}{d t}\right) r \cos\phi \, e^1\wedge e^3 +d\left( \frac{ \kappa a r }{\gamma } \cos\phi\, e^2-\frac{ \kappa a r }{\gamma } \cos\theta \sin\phi \, e^3\right), \label{FLRW-UH-5} \\
		J_H &=& d U_H = J_P = J_{\zeta_2}- J_{\tau \left ( \zeta_2 \right )}\nonumber \\
		& = & -2 \kappa \frac{ r }{a} \cos\theta \sin\phi \left[2 k+2 \left(\frac{d a}{d t}\right)^2+a \frac{d^2 a}{ d t^2}\right] e^0\wedge e^1\wedge e^2 \nonumber\\
		& & -2 \kappa \frac{ r }{a} \cos\phi \left[2 k+2 \left(\frac{d a}{d t}\right)^2 + a \frac{d^2 a}{ d t^2}\right] e^0\wedge e^1\wedge e^3.
	\end{eqnarray}
	Notice that $U_P$ is given by the first five terms in the last equality in \eqref{FLRW-UH-5}.\\
	\item[(vi)] For $\zeta_3$
	\begin{eqnarray}
		U_H &=& U_{\zeta_3} \nonumber\\
		&=&U_P+d\left[ \frac{\kappa }{\gamma} ( \zeta_3 \intprod e_I) e^I \right] \nonumber\\
		&=& -2 \kappa \cos\theta \, e^0\wedge e^1 +2 \kappa \sqrt{1-k r^2} \sin\theta \, e^0\wedge e^2\nonumber\\
		&&+ 2\kappa \left(\frac{d a}{d t}\right) r \sin\theta \, e^1\wedge e^2 + d\left( \frac{ \kappa a r }{\gamma } \sin\theta \, e^3 \right), \label{FLRW-UH-6}\\
		J_H &=& d U_H = J_P = J_{\zeta_3}\nonumber \\
		& = & 2 \kappa \frac{ r}{a} \sin\theta \left[2k+ 2\left(\frac{d a}{d t}\right)^2+a \frac{d^2 a}{ d t^2}\right] e^0\wedge e^1\wedge e^2.
	\end{eqnarray}
	Notice that $U_P$ is given by the first three terms in the last equality in \eqref{FLRW-UH-6}.
\end{enumerate}

\section{Conclusions}\label{Sec:Conclusion}
In this paper,  we   define  off-shell Noether currents and potentials for the $n$-dimensional Palatini Lagrangian and the four-dimensional Holst Lagrangian, which embody first-order formulations of general relativity with a cosmological constant. To derive them, we   implement  a new theoretical framework that uses off-shell Noether identities satisfied by the variational derivatives of each formulation, which,  combined with the variation of the Lagrangian under the infinitesimal versions of the underlying gauge symmetries, lead to the appropriate identification of these off-shell Noether currents and potentials. Two remarkable aspects of our framework are that the whole procedure is carried out {\it off-shell} and that the resulting Noether currents are off-shell conserved too. More precisely, for the $n$-dimensional Palatini Lagrangian, we   derive   off-shell expressions for the Noether currents and potentials associated to diffeomorphisms generated by arbitrary vector fields and local $SO(n-1,1)$ or $SO(n)$ transformations. The resulting off-shell Noether current and potential associated to diffeomorphisms can be regarded as the first-order version of those reported in \cite{Kim1} for general relativity in the metric second-order formalism. In the case of the Holst Lagrangian, the off-shell Noether currents and potentials, for both diffeomorphisms and local $SO(3,1)$ or $SO(4)$ transformations, are affected by the Immirzi parameter in a non-trivial way. Similar   to the $n$-dimensional Palatini Lagrangian, the resulting off-shell Noether current and potential associated to diffeomorphisms can also be regarded   as a first-order version of those reported in \cite{Kim1} for general relativity in the metric second-order formalism. In addition, we   compute  the associated off-shell currents for the so called `improved diffeomorphisms' and for the `generalization of local translations' reported in   \cite{Montesinos1}, showing that they identically vanish for both first-order formulations of general relativity. However, we   also  show that the off-shell Noether current and potential associated to diffeomorphisms emerge from these symmetries. 

For both the $n$-dimensional Palatini and Holst Lagrangians, we   also study  how these off-shell Noether currents and potentials simplify in a spacetime with symmetries generated by Killing vector fields. In particular, for the $n$-dimensional Palatini Lagrangian, we     show   that the action of a Killing vector field on the orthonormal frame and the connection equals a local $SO(n-1,1)$ or $SO(n)$ transformation plus a trivial gauge transformation that only affects the infinitesimal transformation of the connection. The resulting off-shell Noether currents and potentials for this effective gauge transformation are  also   reported,  and they can be expressed, respectively, as the difference of the off-shell Noether currents and potentials associated to Killing vectors and their induced $SO(n)$ or $SO(n-1,1)$ transformations. Analogous results follow from the Holst Lagrangian.

To simplify things a bit, we   consider  the `half off-shell' case, which is defined by the conditions ${\mathcal E} \neq 0$ and ${\mathcal E}_{IJ}=0$ (thus,    we work on solutions of the equation of motion for the connection) for both formulations of general relativity and thus the aforementioned trivial transformation of the connection is set to zero. We    show  that the `half off-shell' Noether currents and potentials for diffeomorphisms and local $SO(3,1)$ or $SO(4)$ transformations for the Holst Lagrangian generically depend on the Immirzi parameter, which is also true in the `on-shell' case. This result is remarkable, since such a contribution is not expected from the point of view of the second-order formalism for general relativity in terms of the tetrad, which is what the Holst Lagrangian collapses to when the condition ${\mathcal E}_{IJ}=0$ is satisfied and does not depend on the Immirzi parameter whatsoever. Furthermore, in the `half off-shell' case, the Noether potential associated to the effective gauge transformation for the Holst Lagrangian differs from that for the Palatini Lagrangian by an exact differential form depending on the Immirzi parameter. To illustrate our approach, we   explicitly compute  the `half off-shell' Noether currents and potentials discussed above, for Killing vector fields, their induced local $SO(3,1)$ transformations, and the associated effective gauge transformations, in four-dimensional static spherically symmetric and FLRW spacetimes, for both Palatini and Holst Lagrangians. For the Holst Lagrangian, the resulting Noether currents and potentials generically depend on the Immirzi parameter, except for the Noether current associated to the effective gauge transformation. 

Although we do  not consider adding boundary terms to the Lagrangians in this paper, they can be handled with our theoretical techniques,  and we expect the addition of boundary terms to the action principles defined by the Palatini and Holst Lagrangians generically contribute to the off-shell Noether currents and their associated potentials. The understanding of such terms in gravity is essential to appropriately define quantities such as asymptotic charges and black hole entropy, and will be one of the main focuses of our forthcoming studies. In addition, those studies might help to clarify the role of the Immirzi parameter in the definition of conserved charges and entropy as well. We expect to confront our results with those obtained in the literature following alternative approaches within the first-order formalism \cite{adami2016,Jacobson,Corichi2,Chakra,Frodden_2018,ADAMI20191,oliveri2020,Barnich_20204D,Freidel1,Freidel2}.

Even though we   construct  the off-shell Noether currents and potentials for general relativity in the first-order formalism, it is obvious that the same theoretical framework can be extended to any gauge theory and, in particular, to any diffeomorphism invariant theory of gravity in the first-order formalism. In particular, similar off-shell Noether currents and potentials can be obtained using the formalism developed in this paper for $f({\mathcal R})$ theories \cite{Montesinos4}, matter fields coupled to general relativity \cite{Montesinos2}, and any other alternative theory of gravity such as  Lovelock gravity \cite{Montesinos3} in the first-order formalism. Moreover, it would also be interesting to study other gravitational models within the first-order formalism including some background structure into play, such as unimodular gravity \cite{Corral,Corral_2019} and extensions thereof. 

\acknowledgments

We thank Alejandro Corichi and Ulises Nucamendi for very valuable comments on the issues addressed in this paper. Mariano Celada gratefully thanks DGAPA-UNAM for financial support. This work was partially supported by Fondo SEP-Cinvestav and by Consejo Nacional de Ciencia y Tecnolog\'{i}a (CONACyT), M\'{e}xico, Grant No. A1-S-7701.

 	\bibliographystyle{apsrev4-1}

\end{document}